\documentclass[12pt]{article}
\usepackage{amsfonts}
\usepackage{amsmath}
\usepackage{amssymb}
\usepackage{graphicx}
\usepackage{color}
\usepackage[all, knot]{xy}
\usepackage{tikz}

\usepackage{ulem}
\usepackage{hyperref}

\usepackage[utf8]{inputenc}
\usepackage{epstopdf}
\usepackage[footnotesize]{caption}
\usepackage{amsthm}
\usepackage{enumitem}
\usepackage{mathrsfs}

\usepackage[margin=3cm]{geometry}

\def \be {\begin{equation}}
\def \ee {\end{equation}}
\def \bea {\begin{eqnarray}}
\def \eea {\end{eqnarray}}
\def \nn {\nonumber}

\def \rr {\raise.35ex\hbox{\small $\prime$}\kern-.17em{\mbox{\large $\imath$}}}

\def \dels {\partial\kern-.6em /\kern.1em}
\def \As {{A\kern-.5em / \kern.5em}}
\def \Ds {D\kern-.7em / \kern.5em}

\def \ks {k\kern-.5em /}
\def \ls {l\kern-.5em /}

\newcommand{\ci}[1]{}

\newcommand{\ba}{\begin{eqnarray}}
\newcommand{\ea}{\end{eqnarray}}
\newcommand{\bal}{\begin{align}}
\newcommand{\eal}{\end{align}}
\newcommand{\bay}[1]{\left(\begin{array}{#1}}
\newcommand{\eay}{\end{array}\right)}

\newcommand{\hide}[1]{}

\DeclareMathOperator{\sech}{sech}

\newlist{axioms}{enumerate}{2}
\setlist[axioms,1]{label=\textbf{A\arabic{axiomsi}.}, ref=A\arabic{axiomsi}}
\setlist[axioms,2]{label=\textbf{A\arabic{axiomsi}\rlap{\myEnumCounter{axiomsii}}.},%
                   ref=A\arabic{axiomsi}\myEnumCounter{axiomsii},%
                   align=parleft,%
                   leftmargin=0em,%
                   itemsep=1.4ex,%
                   before={\stepcounter{axiomsi}}}

  \usetikzlibrary{decorations.markings}

\begin{document}

\begin{titlepage}
\begin{center}

\textbf{\LARGE
AdS$_3$ Einstein Gravity and\\ 
Boundary Description: Pedagogical Review
\vskip.3cm
}
\vskip .5in
{\large
Chen-Te Ma$^{a,b,c,d,e,f}$ \footnote{e-mail address: yefgst@gmail.com}
\\
\vskip 1mm
}
{\sl
$^a$ 
Department of Physics and Astronomy, Iowa State University, Ames, Iowa 50011, US. 
\\
$^b$
Asia Pacific Center for Theoretical Physics,\\
Pohang University of Science and Technology, 
Pohang 37673, Gyeongsangbuk-do, South Korea. 
\\
$^c$
Guangdong Provincial Key Laboratory of Nuclear Science,\\
 Institute of Quantum Matter,
South China Normal University, Guangzhou 510006, Guangdong, China.
\\
$^d$
School of Physics and Telecommunication Engineering,\\
South China Normal University, Guangzhou 510006, Guangdong, China. 
\\
$^e$
Guangdong-Hong Kong Joint Laboratory of Quantum Matter,\\
 Southern Nuclear Science Computing Center, 
South China Normal University, Guangzhou 510006, China.
\\
$^f$
The Laboratory for Quantum Gravity and Strings,\\
 Department of Mathematics and Applied Mathematics,\\
University of Cape Town, Private Bag, Rondebosch 7700, South Africa.
}\\
\vskip 1mm
\vspace{40pt}
\end{center}

\newpage
\begin{abstract} 
We review the various aspects of the 3D Einstein gravity theory with a negative cosmological constant and its boundary description. 
We also explore its connections to CFTs, modular symmetry, and holography. 
It is worth noting that this particular theory is topological in nature, which means that all the physical degrees of freedom are located on the boundary. 
Additionally, we can derive the boundary description on a torus, which takes the form of a 2D Schwarzian theory. 
This observation suggests that the relevant degrees of freedom for the theory can be described using this 2D theory.
Because of the renormalizability of the 3D gravity theory, one can probe the quantum regime. 
This suggests that it is possible to investigate quantum phenomena. 
Unlike the conventional CFTs, when considering the AdS$_3$ background, the boundary theory loses modular symmetry. 
This represents a departure from the usual behavior of CFT and is quite intriguing. 
The Weyl transformation induces anomaly in CFTs, and we indicate that applying this transformation to the 2D Schwarzian theory leads to similar results. 
Summing over all geometries with the asymptotic AdS$_3$ boundary condition is equivalent to summing over a modular group. 
The partition function is one-loop exact and therefore an analytical expression from the summation. 
This theory holds potential applications in Quantum Information and is a recurring theme in the study of holography, where gravitational theories are connected with CFTs. 
\end{abstract}
\end{titlepage}

\section{Introduction}
\label{sec:1}
\noindent 
Black hole entropy \cite{Bekenstein:1973ur} is a concept that assigns a measure of entropy to a black hole. 
This idea emerged from efforts to satisfy the laws of thermodynamics \cite{Bardeen:1973gs} with the physics of black holes. 
The black hole's entropy depends on its mass, electric charge, and angular momentum. 
These parameters are crucial in defining a black hole's properties. 
The entropy of a black hole is related to the area of its event horizon, indicating that entropy should be a monotonic function of the event horizon \cite{Hawking:1974sw}. 
Black hole entropy has profound implications in theoretical physics. 
One of the notable consequences is the {\it Holographic Principle} \cite{tHooft:1993dmi,Susskind:1994vu}. 
The Holographic Principle is a concept in theoretical physics that suggests that the information and degrees of freedom of Quantum Gravity can be encoded or represented on a lower-dimensional boundary surrounding it \cite{tHooft:1993dmi,Susskind:1994vu}. 
\\

\noindent 
The realization of the holographic principle is motivated by String Theory, which describes the dynamics of one-dimensional objects (strings). 
The particular example is a duality in theoretical physics that connects two seemingly different theories: the gravitational theory in a ($d+1$)-dimensional Anti-de Sitter (AdS$_{d+1}$) space and a $d$-dimensional conformal field theory (CFT$_d$). 
This duality establishes a connection between gravity and field theories called {\it AdS/CFT} correspondence. 
String Theory is a candidate for a theory of Quantum Gravity. 
It has the potential to naturally include the graviton, which is the hypothetical quantum particle related to gravity. 
Additionally, it is considered ultraviolet (UV) complete, making it promising evidence for the AdS/CFT correspondence conjecture. 
Einstein gravity in four dimensions is not renormalizable, which is why the AdS/CFT correspondence is valuable. 
However, the issue of renormalizability can be overcome in the context of Quantum Field Theory (QFT), which is another reason why the AdS/CFT correspondence is important. 
As a result, researchers can use the AdS/CFT correspondence to examine gravity within the framework of QFT, providing a distinct perspective on Quantum Gravity. 
Recently, Quantum Information goes into the AdS/CFT correspondence beginning from the Ryu-Takayanagi (RT) conjecture \cite{Ryu:2006bv, Ryu:2006ef}, which connects the entanglement entropy (EE) of CFT$_d$ to the minimum area of surfaces (co-dimension two surfaces on a given time slice) in AdS$_{d+1}$ space \cite{Ryu:2006bv, Ryu:2006ef}. 
This connection suggests that {\it Quantum Entanglement} plays a role in generating spacetime, supporting the idea of {\it Emergent Spacetime}. 
For a spherical entangling surface, a conformal transformation maps EE to thermal entropy \cite{Casini:2011kv}. 
Consistent results were obtained between the holographic method (minimum surface) and field theory. 
The {\it replica trick} is a mathematical technique used to calculate EE in field theory \cite{Holzhey:1994we}. 
When applied to bulk gravity theory, the result also gives the co-dimensional two surface \cite{Lewkowycz:2013nqa,Ma:2016deg,Ma:2016pah}. 
Introducing the Hayward term \cite{Hayward:1993my} to the gravity theory is an equivalent approach to avoiding the use of the replica trick at the classical level \cite{Takayanagi:2019tvn}. 
Rényi entropy also has a similar holographic formulation through the cosmic brane \cite{Dong:2016fnf} and the Hayward term \cite{Botta-Cantcheff:2020ywu}. 
Hence, the minimum surface approach provides various pieces of evidence at the classical level. 
The concept of excitation entanglement entropy is introduced to illustrate the difference between the entanglement entropy of general excitation and their arbitrary descendants \cite{Sheikh-Jabbari:2016znt}. 
This observable is a finite quantity independent of the cutoff and can be computed by the holographic formula \cite{Sheikh-Jabbari:2016znt}. 
The holographic formula is a tool that simplifies the study of many-body physics by reducing the complexity of computations related to EE. 
Because we hope to have the gauge invariant measure, the bipartition becomes subtle if a holographic system has the gauge symmetry. 
However, from the operator algebra point of view, we can define gauge invariant measures with consistent results in the $p$-form gauge theory \cite{Ma:2015xes,Huang:2016bkp}. 
In AdS$_3$/CFT$_2$ case, the RT conjecture can have the simplified interpretation of the relationship between the length of the geodesic line and the entanglement entropy by the differential entropy \cite{Balasubramanian:2013lsa,Czech:2014ppa} or kinematic space \cite{Czech:2015qta,Czech:2015kbp}.   
\\

\noindent 
Because it is hard to have the complete Lagrangian description for the bulk gravitational theory and CFT simultaneously, we only have a few examples to establish concrete AdS/CFT correspondence by comparing results between the bulk and boundary theories. 
One of the most well-known examples of the AdS/CFT correspondence involves String Theory in AdS$_5\times S^5$, where $S^5$ is a five-dimensional sphere manifold, is dual to ${\cal N}=4$ Super Yang-Mills (SYM) theory in four dimensions. 
In String Theory, there is a clear correspondence between perturbation computations and integrability techniques in SYM theory. 
However, we do not have a direct way to derive SYM theory from String Theory. 
In the realm of 3D Einstein gravity, there are no propagating gravitational degrees of freedom in the bulk. 
Instead, the physical degrees of freedom reside on the boundary. 
This allows for direct derivation of the boundary theory. 
Moreover, the system is simple enough to compute the stress tensor $n$-point correlators with ease \cite{He:2023hoj}. 
The most general solutions to the AdS$_3$ Einstein gravity with Brown-Henneaux boundary conditions can also be classified \cite{Sheikh-Jabbari:2016unm,Sheikh-Jabbari:2016npa,OColgain:2016msw}. 
The doubled Chern-Simons theory with SL(2) gauge groups reformulates the 3D Einstein gravity theory and resolves the issue of renormalizability with a dimensionless coupling constant \cite{Witten:1988hc}. 
Because the functional measure of the metric formulation is only over the non-singular vielbeins, the metric formulation differs from the gauge formulation non-perturbatively \cite{Witten:2007kt}. 
However, the Chern-Simons description \cite{Elitzur:1989nr} provides a simple way to treat the quantum correction of 3D Einstein gravity. 
When considering a negative cosmological constant in 3D Einstein gravity, the conformal symmetry corresponds to the bulk gauge symmetry (or SO(2, 2) gauge group). 
According to the theory, the boundary is expected to be a CFT$_2$ \cite{Brown:1986nw}. 
However, research has found that pure 3D Einstein gravity does not possess a physical CFT description \cite{Maloney:2007ud,Manschot:2007ha}. 
The summation over the asymptotic AdS$_3$ boundary condition results in a negative density of states \cite{Maloney:2007ud,Manschot:2007ha}. 
This negative density is associated with a non-unitary CFT. 
The error approximating the discrete spectrum to a continuous spectrum that is not sensitive to the spin \cite{Mukhametzhanov:2019pzy,Ganguly:2019ksp} and the generalization of the usual Cardy formula \cite{Kusuki:2019gjs} help the CFT computation to support the same conclusion \cite{Benjamin:2019stq,Maxfield:2019hdt}. 
The boundary theory of AdS$_3$ Einstein gravity has been a subject of study. 
It was initially proposed as Liouville theory \cite{Coussaert:1995zp, Rooman:2000zi}, but this proposal was not acceptable due to issues with the non-normalizable vacuum. 
Recent work has shown that the correct boundary on the torus manifold (asymptotic boundary for the Euclidean case) has SL(2) gauge symmetry with the Schwarzian form \cite{Cotler:2018zff}, leading to what is called {\it 2D Schwarzian theory}. 
This theory is dual to chiral scalar fields \cite{Huang:2020tjl,Floreanini:1987as,Tseytlin:1990ar}, and the classical limit is the Liouville theory \cite{Nguyen:2021pdz}. 
\\

\noindent 
The partition function of 2D Schwarzian theory on the torus manifold is {\it one-loop exact} \cite{Cotler:2018zff,Giombi:2008vd}. 
This suggests that an analytical solution for the partition function can be obtained without considering higher-loop contributions. 
It can be generalized to an $n$-sheet covering and related to EE with the bulk dual, known as the Wilson line \cite{Ammon:2013hba, deBoer:2013vca, Huang:2019nfm}. 
Even though 2D Schwarzian theory lacks a covariant Lagrangian description, the {\it Weyl transformation} acting on the torus case generates the {\it Liouville theory} \cite{Huang:2023aqz}. 
This allows for simplifications using Liouville theory or Weyl anomaly, similar to CFT$_2$ \cite{Huang:2023aqz, Headrick:2010zt}. 
The 2D Schwarzian theory is a simple quantum system for studying the bulk/boundary correspondence.   
When considering the AdS/CFT correspondence on a wormhole manifold, introducing an ensemble to the boundary theory is a common approach \cite{Cotler:2021cqa}. 
However, this ensemble averaging does not provide a smooth description of $N$ to the Hamiltonian $H_N$ describing a black hole state \cite{Schlenker:2022dyo}, which implies differences between the boundary dual of Einstein gravity and String Theory in this context \cite{Schlenker:2022dyo,Chandra:2022bqq}.  

\subsection{Outline}
\noindent
The outline of this review is as follows. 
This section introduces the gauge formulation of 3D Einstein Gravity in Sec.~\ref{sec:2}. 
It covers the basic concepts and equations related to this topic. 
\\

\noindent 
We explore the boundary description of AdS$_3$ Einstein Gravity in Sec.~\ref{sec:3}. 
It includes discussions on the 2D Schwarizna theory, the equivalence between chiral scalar fields and the boundary theory, and the examination of the Weyl transformation.  
Additionally, we will calculate the partition function by summing over all manifolds with the asymptotic AdS$_3$ boundary condition.   
\\

\noindent 
We introduce the concept of holographic EE in Sec.~\ref{sec:4}. 
We plan to demonstrate the RT conjecture using the AdS$_3$ geometry or an exact solution. 
Furthermore, we will apply the gauge formulation to find the dual of EE using the bulk Wilson line. 
We discuss potential future research directions or areas of study related to the topics in Sec.~\ref{sec:5}.  

\section{3D Einstein Gravity}
\label{sec:2}
\noindent
We first introduce the second-order and first-order formulations of 3D Einstein Gravity and then rewrite it to get the gauge formulation, SL(2) doubled Chern-Simons \cite{Witten:2007kt}. 
Because the coupling constant in the gauge formulation is dimensionless, it is easier to treat the quantum fluctuation. 
The path integration is over the singular and non-singular vielbeins. 
We can find the difference at the non-perturbative level. 

\subsection{Second-Order Formulation} 
\noindent 
The Lagrangian description of metric formulation (or second-order formulation) for 3D Einstein Gravity with the Lorentz signature $(-1, 1, 1)$ is 
\bea
S_{3DS}=\frac{1}{16\pi G_3}\int d^3x\ \sqrt{|\det -g_{\mu\nu}|}(R-2\Lambda), 
\label{3DS}
\eea
where $G_3$ is the 3D gravitational constant, and $\Lambda$ is the cosmological constant. 
The $g_{\mu\nu}$ represents the metric that is used to define the spacetime interval. 
This interval can be determined by the formula 
\bea
ds^2=g_{\mu\nu}dx^{\mu}dx^{\nu}.  
\eea 
We denote the bulk spacetime indices by $\mu, \nu, \cdots$. 
The Ricci curvature tensor $R_{\mu\nu}$ is defined by
\bea
R_{\mu\nu}\equiv\partial_{\delta}\Gamma^{\delta}_{\nu\mu}-\partial_{\nu}\Gamma^{\delta}_{\delta\mu}
+\Gamma^{\delta}_{\delta\lambda}\Gamma^{\lambda}_{\nu\mu}
-\Gamma^{\delta}_{\nu\lambda}\Gamma^{\lambda}_{\delta\mu}, 
\eea
where Christoffel symbol $\Gamma^{\mu}_{\nu\delta}$ is 
\bea
 \Gamma^{\mu}_{\nu\delta}&\equiv&\frac{1}{2}g^{\mu\lambda}\bigg(\partial_{\delta}g_{\lambda\nu}+\partial_{\nu}g_{\lambda\delta}
-\partial_{\lambda}g_{\nu\delta}\bigg),
\eea 
The Ricci scalar is a mathematical object defined using the metric tensor and the Ricci tensor. 
It is given by the expression
\bea
 R\equiv g^{\mu\nu}R_{\mu\nu}, 
\eea 
where $g^{\mu\nu}$ is the inverse of the metric tensor and $R_{\mu\nu}$ is the Ricci tensor. 

\subsection{First-Order Formulation}
\noindent
The action \eqref{3DS} can be rewritten as the first-order formulation from the vielbein 
\bea
e_{\mu}\equiv e_{\mu}{}^aJ_a
\eea
and spin connection 
\bea
\omega_{\mu}\equiv\omega_{\mu}{}^aJ_a
\eea
\bea
S_{3DF}=\frac{1}{16\pi G_3}\int d^3x\ e(R_{\mu\nu}{}^{ab}e^{\mu}{}_ae^{\nu}{}_b-2\Lambda), 
\eea
where 
\bea
R_{\mu\nu}{}^{ab}&=&\partial_{\mu}\omega_{\nu}{}^{ab}-\partial_{\nu}\omega_{\mu}{}^{ab}
+(\omega_{\mu}{}^{ac}\omega_{\nu}{}^{bd}-\omega_{\nu}{}^{ac}\omega_{\mu}{}^{bd})\eta_{cd};  
\nn\\
e_d&\equiv&\sqrt{\det(-e_{\mu, a}e_{\nu}^a)}.  
\eea 
The $\omega_{\mu}{}^{bc}$ is defined by the spin connection
\bea
\omega_{\mu}{}^a=\frac{1}{2}\epsilon^a{}_{bc}\omega_{\mu}{}^{bc}. 
\eea
The $J_a$ are SL(2) generators:
\bea
J_0\equiv\begin{pmatrix}
0&-\frac{1}{2}
\\
\frac{1}{2}& 0
\end{pmatrix}; \ 
J_1\equiv\begin{pmatrix}
0&\frac{1}{2}
\\
\frac{1}{2}&0
\end{pmatrix}; \ 
J_2\equiv\begin{pmatrix}
\frac{1}{2}& 0
\\
0&-\frac{1}{2}
\end{pmatrix}, 
\eea 
which satisfies the commutation and the trace relations: 
\bea
\lbrack J^a, J^b\rbrack=\epsilon^{abc}; \ \mathrm{Tr}(J^a J^b)=\frac{1}{2}\eta^{ab}. 
\eea
The Lie algebra indices are raised or lowered by 
\bea
\eta\equiv\mathrm{diag}(-1, 1, 1). 
\eea 
The metric now can be rewritten in terms of vielbein as
\bea
g_{\mu\nu}=2\mathrm{Tr}(e_{\mu}e_{\nu}). 
\label{MV}
\eea
We will reformulate 3D Einstein Gravity via the gauge fields (gauge formulation) \cite{Witten:1988hc}. 

\subsection{Gauge Formulation}
\noindent 
We reformulate 3D Einstein Gravity with a negative cosmological constant from the Chern-Simons theory \cite{Witten:1988hc}
\bea
S_{3DG}=S_{\mathrm{CS}}(A)-S_{\mathrm{CS}}(\bar{A}), 
\eea
where 
\bea
S_{\mathrm{CS}}=\frac{k}{4\pi}\int d^3 x\ \mbox{Tr}\bigg(\epsilon^{\mu\nu\rho}A_{\mu}\partial_{\nu}A_{\rho}+\frac{2}{3}\epsilon^{\mu\nu\rho}A_{\mu}A_{\nu}A_{\rho}\bigg),
\eea
The constant $k$ is defined as \cite{Witten:1988hc}
\bea
k\equiv\frac{l}{4G_3}, 
\eea
where 
\bea
\frac{1}{l^2}\equiv-\Lambda. 
\eea
The gauge fields are defined by the vielbein and spin connection \cite{Witten:1988hc}:
\bea
A_{\mu}=J_a\bigg(\omega_{\mu}{}^a+\frac{1}{l}e_{\mu}{}^a\bigg), \qquad
\bar{A}_{\mu}=J_a\bigg(\omega_{\mu}{}^a-\frac{1}{l}e_{\mu}{}^a\bigg). 
\label{GVS}
\eea
We substitute the vielbein and spin connection to the gauge fields, and then we obtain that: 
\bea
&&
\epsilon^{\mu\nu\rho}A_{\mu}\partial_{\nu}A_{\rho}
\nn\\
&=&\epsilon^{\mu\nu\rho}J_aJ_b\bigg(
\omega_{\mu}{}^a\partial_{\nu}\omega_{\rho}{}^b+\frac{1}{l}\omega_{\mu}{}^a\partial_{\nu}e_{\rho}{}^b+\frac{1}{l}e_{\mu}{}^a\partial_{\nu}\omega_{\rho}{}^b+\frac{1}{l^2}e_{\mu}{}^a\partial_{\nu}e_{\rho}{}^b\bigg);
\nn\\
&&
\epsilon^{\mu\nu\rho}\bar{A}_{\mu}\partial_{\nu}\bar{A}_{\rho}
\nn\\
&=&\epsilon^{\mu\nu\rho}J_aJ_b\bigg(
\omega_{\mu}{}^a\partial_{\nu}\omega_{\rho}{}^b-\frac{1}{l}\omega_{\mu}{}^a\partial_{\nu}e_{\rho}{}^b-\frac{1}{l}e_{\mu}{}^a\partial_{\nu}\omega_{\rho}{}^b+\frac{1}{l^2}e_{\mu}{}^a\partial_{\nu}e_{\rho}{}^b\bigg);
\nn\\
&&
\frac{2}{3}\epsilon^{\mu\nu\rho}A_{\mu}A_{\nu}A_{\rho}
\nn\\
&=&\frac{2}{3}\epsilon^{\mu\nu\rho}J_aJ_bJ_c\bigg(\omega_{\mu}{}^a\omega_{\nu}{}^b\omega_{\rho}{}^c+\frac{1}{l}\omega_{\mu}{}^a\omega_{\nu}{}^b e_{\rho}{}^c+\frac{1}{l}\omega_{\mu}{}^a e_{\nu}{}^b\omega_{\rho}{}^c+\frac{1}{l}e_{\mu}{}^a\omega_{\nu}{}^b\omega_{\rho}{}^c
\nn\\
&&
+\frac{1}{l^2}\omega_{\mu}{}^a e_{\nu}{}^b e_{\rho}{}^c+\frac{1}{l^2}e_{\mu}{}^a\omega_{\nu}
{}^b e_{\rho}{}^c+\frac{1}{l^3}e_{\mu}{}^a e_{\nu}{}^b e_{\rho}{}^c\bigg);
\nn\\
&&
\frac{2}{3}\epsilon^{\mu\nu\rho}\bar{A}_{\mu}\bar{A}_{\nu}\bar{A}_{\rho}
\nn\\
&=&
\frac{2}{3}\epsilon^{\mu\nu\rho}J_aJ_bJ_c\bigg(\omega_{\mu}{}^a\omega_{\nu}{}^b\omega_{\rho}{}^c-\frac{1}{l}\omega_{\mu}{}^a\omega_{\nu}{}^b e_{\rho}{}^c-\frac{1}{l}\omega_{\mu}{}^a e_{\nu}{}^b\omega_{\rho}{}^c-\frac{1}{l}e_{\mu}{}^a\omega_{\nu}{}^b\omega_{\rho}{}^c
\nn\\
&&
+\frac{1}{l^2}\omega_{\mu}{}^a e_{\nu}{}^b e_{\rho}{}^c+\frac{1}{l^2}e_{\mu}{}^a\omega_{\nu}
{}^b e_{\rho}{}^c-\frac{1}{l^3}e_{\mu}{}^a e_{\nu}{}^b e_{\rho}{}^c\bigg). 
\label{3DG1}
\eea
From Eq. \eqref{3DG1}, we obtain the action: 
\bea
&&
S_{3DG}
\nn\\
&=&S_{\mathrm{CS}}(A)-S_{\mathrm{CS}}(\bar{A})
\nn\\
&=&\frac{k}{4\pi }\int d^3 x\ \epsilon^{\mu\nu\rho} \mbox{Tr}\bigg\lbrack J_aJ_b\bigg(\frac{1}{l}\omega_{\mu}{}^a\partial_{\nu}e_{\rho}{}^b
+\frac{1}{l}e_{\mu}{}^a\partial_{\nu}\omega_{\rho}{}^b\bigg)
\nn\\
&&
+J_aJ_b\bigg(\frac{1}{l}\omega_{\mu}{}^a\partial_{\nu}e_{\rho}{}^b
+\frac{1}{l}e_{\mu}{}^a\partial_{\nu}\omega_{\rho}{}^b\bigg)
\nn\\
&&+\frac{2}{3}J_aJ_bJ_c\bigg(\frac{1}{l}\omega_{\mu}{}^a\omega_{\nu}{}^b e_{\rho}{}^c
+\frac{1}{l}\omega_{\mu}{}^a e_{\nu}{}^b\omega_{\rho}{}^c+e_{\mu}{}^a\omega_{\nu}{}^b\omega_{\rho}{}^c
+\frac{1}{l^3}e_{\mu}{}^a e_{\nu}{}^b e_{\rho}{}^c\bigg)
\nn\\
&&
+\frac{2}{3}J_aJ_bJ_c\bigg(\frac{1}{l}\omega_{\mu}{}^a\omega_{\nu}{}^b e_{\rho}{}^c
+\frac{1}{l}\omega_{\mu}{}^a e_{\nu}{}^b\omega_{\rho}{}^c
+\frac{1}{l}e_{\mu}{}^a\omega_{\nu}{}^b\omega_{\rho}{}^c
+\frac{1}{l^3}e_{\mu}{}^a e_{\nu}{}^b e_{\rho}{}^c\bigg)
\bigg\rbrack.
\nn\\
\label{3DG}
\eea
\\

\noindent
Now we compute the first two terms of the last equality in Eq. \eqref{3DG}:
\bea
&&\frac{k}{4\pi}\int d^3 x\ \mbox{Tr}\bigg\lbrack\epsilon^{\mu\nu\rho}J_aJ_b\bigg(\frac{1}{l}\omega_{\mu}{}^a\partial_{\nu}e_{\rho}{}^b
+\frac{1}{l}e_{\mu}{}^a\partial_{\nu}\omega_{\rho}{}^b\bigg)\bigg\rbrack
\nn\\
&=&\frac{k}{4\pi}\int d^3 x\ \bigg\lbrack\frac{1}{2l}\epsilon^{\mu\nu\rho}\bigg(\partial_{\mu}\omega_{\nu}{}^a -\partial_{\nu}\omega_{\mu}{}^a \bigg)e_{\rho,a}\bigg\rbrack;  
\nn\\
&&-\frac{k}{4\pi}\int d^3 x\ \mbox{Tr}\bigg\lbrack\epsilon^{\mu\nu\rho}J_aJ_b\bigg(\frac{1}{l}\omega_{\mu}{}^a\partial_{\nu}e_{\rho}{}^b
+\frac{1}{l}e_{\mu}{}^a\partial_{\nu}\omega_{\rho}{}^b\bigg)\bigg\rbrack
\nn\\
&=&\frac{k}{4\pi}\int d^3 x\ \bigg\lbrack\frac{1}{2l}\epsilon^{\mu\nu\rho}\bigg(\partial_{\mu}\omega_{\nu}{}^a -\partial_{\nu}\omega_{\mu}{}^a \bigg)e_{\rho, a}\bigg\rbrack.   
\eea
The above equalities are up to a total derivative term. 
Hence we obtain that:
\bea
&&\frac{k}{4\pi}\int d^3 x\ \mbox{Tr}\bigg\lbrack\epsilon^{\mu\nu\rho}J_aJ_b\bigg(\frac{1}{l}\omega_{\mu}{}^a\partial_{\nu}e_{\rho}{}^b
+\frac{1}{l}e_{\mu}{}^a\partial_{\nu}\omega_{\rho}{}^b\bigg)\bigg\rbrack
\nn\\
&&
-\frac{k}{4\pi}\int d^3 x\ \mbox{Tr}\bigg\lbrack\epsilon^{\mu\nu\rho}J_aJ_b\bigg(\frac{1}{l}\omega_{\mu}{}^a\partial_{\nu}e_{\rho}{}^b
+\frac{1}{l}e_{\mu}{}^a\partial_{\nu}\omega_{\rho}{}^b\bigg)\bigg\rbrack
\nn\\
&=&
\frac{k}{4\pi l}\int d^3 x\ \bigg\lbrack\epsilon^{\mu\nu\rho}\bigg(\partial_{\mu}\omega_{\nu}{}^a -\partial_{\nu}\omega_{\mu}{}^a \bigg)e_{\rho, a}\bigg\rbrack.
\eea
Due to the following identity
\bea
\epsilon^{\mu\nu\rho}\epsilon_{abc}=e_d\bigg(e^{\mu}{}_{[a}e^{\nu}{}_be^{\rho}{}_{c]}\bigg),
\eea
we obtain that:
\bea
&&\epsilon^{\mu\nu\rho}\partial_{\mu}\omega_{\nu}{}^a e_{\rho, a}
\nn\\
&=&
\frac{1}{2}\epsilon^{\mu\nu\rho}\epsilon^{abc}\big(\partial_{\mu}\omega_{\nu, bc}\big) e_{\rho, a}
\nn\\
&=&\frac{1}{2}e_d\bigg(e^{\mu}{}_{[a}e^{\nu}{}_be^{\rho}{}_{c]}\bigg)\big(\partial_{\mu}\omega_{\nu}{}^{ bc}\big) e_{\rho}{}^a
\nn\\
&=&e_d\bigg(e^{\mu}{}_{a}e^{\nu}{}_be^{\rho}{}_{c}\bigg)\big(\partial_{\mu}\omega_{\nu}{}^{ bc} \big)e_{\rho}{}^a
+e_d\bigg(e^{\mu}{}_{c}e^{\nu}{}_a e^{\rho}{}_{b}\bigg)\big(\partial_{\mu}\omega_{\nu}{}^{ bc}\big) e_{\rho}{}^a
\nn\\
&&
+e_d\bigg(e^{\mu}{}_{b}e^{\nu}{}_c e^{\rho}{}_{a}\bigg)\big(\partial_{\mu}\omega_{\nu}{}^{ bc}\big) e_{\rho}{}^a
\nn\\
&=&e_d\bigg(e^{\nu}{}_be^{\mu}{}_{c}\bigg)\partial_{\mu}\omega_{\nu}{}^{ bc}
+e_d\bigg(e^{\mu}{}_{c} e^{\nu}{}_{b}\bigg)\partial_{\mu}\omega_{\nu}{}^{ bc} 
+3e_d\bigg(e^{\mu}{}_{b}e^{\nu}{}_c \bigg)\partial_{\mu}\omega_{\nu}{}^{ bc} 
\nn\\
&=&
e_d\bigg(e^{\mu}{}_{b}e^{\nu}{}_c \bigg)\partial_{\mu}\omega_{\nu}{}^{ bc}, 
\eea
in which we used 
\bea
e^{\mu}{}_a e_{\mu}{}^b=\eta^{b}{}_{a}
\eea
in the fourth equality.
Hence we obtain
\bea
&&
\frac{k}{4\pi}\int d^3 x\ \bigg\lbrack\frac{1}{2l}\epsilon^{\mu\nu\rho}\bigg(\partial_{\mu}\omega_{\nu}{}^a -\partial_{\nu}\omega_{\mu}{}^a \bigg)e_{\rho, a}\bigg\rbrack
\nn\\
&=&
\frac{k}{4\pi l}\int d^3x\ e_d\bigg(e^{\mu}{}_{b}e^{\nu}{}_c \bigg)\bigg(\partial_{\mu}\omega_{\nu}^{bc}
-\partial_{\nu}\omega_{\mu}^{bc}\bigg).  
\label{FG1}
\eea
\\

\noindent
We regulate the remaining terms of Eq. \eqref{3DG} as 
\bea
&&\frac{k}{4\pi l }\int d^3 x\ \epsilon^{\mu\nu\rho} \mbox{Tr}\bigg\lbrack
\frac{2}{3}J_aJ_bJ_c\bigg(\omega_{\mu}{}^a\omega_{\nu}{}^b e_{\rho}{}^c+\omega_{\mu}{}^a e_{\nu}{}^b\omega_{\rho}{}^c+e_{\mu}{}^a\omega_{\nu}{}^b\omega_{\rho}{}^c
\nn\\
&&+\frac{1}{l^2}e_{\mu}{}^a e_{\nu}{}^b e_{\rho}{}^c\bigg)
\nn\\
&&
+\frac{2}{3}J_aJ_bJ_c\bigg(\omega_{\mu}{}^a\omega_{\nu}{}^b e_{\rho}{}^c+\omega_{\mu}{}^a e_{\nu}{}^b\omega_{\rho}{}^c+e_{\mu}{}^a\omega_{\nu}{}^b\omega_{\rho}{}^c+\frac{1}{l^2}e_{\mu}{}^a e_{\nu}{}^b e_{\rho}{}^c\bigg)
\bigg\rbrack
\nn\\
&=&\frac{k}{4\pi l}\int d^3x\ \epsilon^{\mu\nu\rho}\mathrm{Tr}\bigg\lbrack
\frac{4}{3}J_aJ_bJ_c\bigg(\omega_{\mu}{}^a\omega_{\nu}{}^b e_{\rho}{}^c+\omega_{\mu}{}^a e_{\nu}{}^b\omega_{\rho}{}^c+e_{\mu}{}^a\omega_{\nu}{}^b\omega_{\rho}{}^c
\nn\\
&&
+\frac{1}{l^2}e_{\mu}{}^a e_{\nu}{}^b e_{\rho}{}^c\bigg)\bigg\rbrack. 
\label{3DG2}
\nn\\
\eea
Now we rewrite the first term of Eq. \eqref{3DG2}: 
\bea
&&\frac{k}{4\pi l}\int d^3 x\ \mathrm{Tr}\bigg(\frac{2}{3}\epsilon^{\mu\nu\rho}J_aJ_bJ_c\omega_{\mu}{}^a\omega_{\nu}{}^b e_{\rho}{}^c-\frac{2}{3}\epsilon^{\mu\nu\rho}J_aJ_cJ_b
\omega_{\mu}{}^a\omega_{\nu}{}^be_{\rho}{}^c\bigg)
\nn\\
&=&\frac{k}{4\pi l}\int d^3 x\ \mathrm{Tr}\bigg(\frac{2}{3}\epsilon^{\mu\nu\rho}J_a\lbrack J_b, J_c\rbrack
\omega_{\mu}{}^a\omega_{\nu}{}^be_{\rho}{}^c\bigg)
\nn\\
&=&\frac{k}{4\pi l}\int d^3 x\ \mathrm{Tr}\bigg(\frac{2}{3}\epsilon^{\mu\nu\rho}\epsilon_{bcd}J_aJ^d
\omega_{\mu}{}^a\omega_{\nu}{}^b e_{\rho}{}^c\bigg)
\nn\\
&=&\frac{k}{4\pi l}\int d^3 x\ \bigg( 
\frac{1}{3}\epsilon^{\mu\nu\rho}\epsilon_{bca}\omega_{\mu}{}^a\omega_{\nu}{}^be_{\rho}{}^c
\bigg),
\eea
then we can get
\bea
&&
\frac{k}{4\pi l}\int d^3x\ \epsilon_{\mu\nu\rho}\mathrm{Tr}\bigg\lbrack
\frac{4}{3}J_aJ_bJ_c\bigg(\omega_{\mu}{}^a\omega_{\nu}{}^b e_{\rho}{}^c+\omega_{\mu}{}^a e_{\nu}{}^b\omega_{\rho}{}^c+e_{\mu}{}^a\omega_{\nu}{}^b\omega_{\rho}{}^c\bigg)\bigg\rbrack
\nn\\
&=&
\frac{k}{4\pi l}\int d^3 x\ 
\epsilon^{\mu\nu\rho}\epsilon_{bca}\omega_{\mu}{}^a\omega_{\nu}{}^be_{\rho}{}^c. 
\label{3DG3}
\eea
By the following equalities: 
\bea
&&\epsilon^{\mu\nu\rho}\epsilon_{bca}\omega_{\mu}{}^a\omega_{\nu}{}^b e_{\rho}{}^c
\nn\\
&=&3e_de^{\mu}{}_ae^{\nu}{}_b \omega_{\mu}{}^a\omega_{\nu}{}^b
-3e_de^{\mu}{}_be^{\nu}{}_a\omega_{\mu}{}^a\omega_{\nu}{}^b
\nn\\
&&+e_de^{\nu}{}_ae^{\mu}{}_b\omega_{\mu}{}^a\omega_{\nu}{}^b
-e_de^{\nu}{}_be^{\mu}{}_a\omega_{\mu}{}^a\omega_{\nu}{}^b
\nn\\
&&
+e_de^{\mu}{}_be^{\nu}{}_a\omega_{\mu}{}^a\omega_{\nu}{}^b
-e_de^{\mu}{}_ae^{\nu}{}_b\omega_{\mu}{}^a\omega_{\nu}{}^b
\nn\\
&=&-\frac{1}{4}e_de^{\mu}{}_b e^{\nu}{}_a \epsilon^a{}_{cd}\epsilon^b{}_{ef}\omega_{\mu}{}^{cd}\omega_{\nu}{}^{ef}
+\frac{1}{4}e_de^{\mu}{}_ae^{\nu}{}_b\epsilon^a{}_{cd}\epsilon^b{}_{ef}\omega_{\mu}{}^{cd}\omega_{\nu}{}^{ef}
\nn\\
&=&-e_de^{\mu}{}_be^{\nu}{}_a f^{ab}{}_{cd\ ef}\omega_{\mu}{}^{cd}\omega_{\nu}{}^{ef},
\eea
where
\bea
f^{ab}{}_{cd\ ef}=\frac{1}{4}\bigg(\epsilon^{a}{}_{cd}\epsilon^b{}_{ef}-\epsilon^b{}_{cd}\epsilon^a{}_{ef}\bigg),
\eea
we can obtain
\bea
&&\frac{k}{4\pi l}\int d^3 x\ 
\epsilon^{\mu\nu\rho}\epsilon_{bca}\omega_{\mu}{}^a\omega_{\nu}{}^be_{\rho}{}^c
\nn\\
&=&-\frac{k}{4\pi l}\int d^3x\
e_de^{\mu}{}_ae^{\nu}{}_b\bigg(\omega_{\mu}{}^{ac}\omega_{\nu}{}^{db}-\omega_{\nu}{}^{ac}\omega_{\mu}{}^{db}\bigg)\eta_{cd}. 
\label{FG2}
\eea
The last term of Eq. \eqref{3DG2} can be rewritten as the cosmological constant term: 
\bea
&&\frac{k}{4\pi l}\int d^3 x\ \mbox{Tr}\bigg(\frac{4}{3l^2}\epsilon^{\mu\nu\rho}J_aJ_bJ_c e_{\mu}{}^ae_{\nu}{}^be_{\rho}{}^c\bigg)
\nn\\
&=&\frac{k}{4\pi l}\int d^3 x\ \mbox{Tr}\bigg(\frac{2}{3l^2}\epsilon^{\mu\nu\rho}\epsilon_{abd}J^dJ_ce_{\mu}{}^ae_{\nu}{}^be_{\rho}{}^c\bigg)
\nn\\
&=&\frac{k}{4\pi l}\int d^3 x\ \bigg(\frac{1}{3l^2}\epsilon^{\mu\nu\rho}\epsilon_{abc}e_{\mu}{}^ae_{\nu}{}^be_{\rho}{}^c\bigg)
\nn\\
&=&\frac{k}{4\pi l}\int d^3x\ e_d\frac{2}{l^2}, 
\label{FG3}
\eea
in which we used:
\bea
&&\mathrm{Tr}\big(\epsilon^{\mu\nu\rho}J_aJ_bJ_ce_{\mu}{}^ae_{\nu}{}^be_{\rho}{}^c\big)
\nn\\
&=&-\mathrm{Tr}\big(\epsilon^{\mu\nu\rho}J_bJ_aJ_ce_{\mu}{}^ae_{\nu}{}^be_{\rho}{}^c\big)
\nn\\
&=&\frac{1}{2}\epsilon^{\mu\nu\rho}\mathrm{Tr}\big(\lbrack J_a, J_b\rbrack J_ce_{\mu}{}^ae_{\nu}{}^be_{\rho}{}^c\big)
\nn\\
&=&\frac{1}{2}\epsilon^{\mu\nu\rho}\epsilon_{abd}\mathrm{Tr}\big(J^dJ_ce_{\mu}{}^ae_{\nu}{}^be_{\rho}{}^c\big)
\eea
in the first equality. 
By combining Eqs. \eqref{FG1}, \eqref{FG2}, \eqref{FG3}, we show the equivalence between the gauge formulation and the first-order formulation \cite{Witten:1988hc}
\bea
S_{3DG}=S_{3DF},  
\eea
up to a total derivative term. 
Because $G_3$ is not a dimensionless constant but $k$ is, the gauge formulation is more convenient to treat the quantum fluctuation \cite{Witten:1988hc}. 
The path integration is only over the non-singular vielbein in the metric formulation
\bea
\int{\cal D}g_{\mu\nu},
\eea
but the measure of the gauge formulation 
\bea
\int{\cal D}A{\cal D}\bar{A}
\eea
is also over the singular one. 
Therefore, the metric and gauge formulations are distinct from the non-perturbative effect, as stated in Ref. \cite{Witten:2007kt}. 

\section{Boundary Description}
\label{sec:3} 
\noindent 
Because the gauge formulation has the SL(2)$\times$SL(2) gauge symmetry \cite{Witten:2007kt}, the boundary theory has the conformal symmetry. 
Therefore, we first review the conformal symmetry. 
We then derive the boundary description of AdS$_3$ Einstein gravity theory, 2D Schwarzian theory \cite{Cotler:2018zff}, which is also dual to chiral scalar fields \cite{Huang:2020tjl}, and discuss the loss of modular symmetry \cite{Cotler:2018zff}. 
We cannot apply the CFT$_2$ result of the Weyl anomaly directly to the 2D Schwarzian theory due to covariance loss in the Lagrangian \cite{Cotler:2018zff}. 
Nevertheless, we perform a direct calculation on the torus manifold to demonstrate the Liouville theory resulting from the Weyl transformation \cite{Huang:2023aqz}. 
Finally, we present the analytical expression for the partition function when summing over all manifolds with the asymptotic AdS$_3$ boundary condition \cite{Maloney:2007ud}.  

\subsection{CFT}
\noindent 
We first introduce the conformal transformation and then the conformal algebra in CFT$_d$. 
We then discuss the correspondence between the gauge symmetry of the gauge formulation and the conformal symmetry. 
Because the gauge symmetry is preserved even with quantum correction, the boundary description of gauge formulation is from CFT$_2$. 

\subsubsection{Conformal Transformation} 
\noindent 
A conformal transformation is an invertible coordinate transformation 
\bea
x^{\mu}\rightarrow \tilde{x}^{\mu}
\eea 
with a transformation of a metric field 
\bea
\tilde{g}_{\mu\nu}(\tilde{x})=\Omega(x)g_{\mu\nu}(x). 
\eea
\\ 

\noindent 
When $d\neq 2$, a conformal transformation of the field $\phi$ is 
\bea
\delta\phi(z)=\epsilon\partial\phi(z)+\Delta(\partial\epsilon)\phi(z), 
\eea 
where 
\bea
\partial\equiv\frac{\partial}{\partial z}. 
\eea
The $\epsilon$ is a holomorphic transformation parameter, and $\Delta$ is a conformal dimension of the holomorphic part. 
When $d=2$, a conformal transformation of the field $\phi$ needs to include the anti-holomorphic part 
\bea
\delta\phi(z, \bar{z})=\epsilon\partial\phi(z, \bar{z})+\Delta(\partial\epsilon)\phi(z, \bar{z})+\bar{\epsilon}\bar{\partial}\phi(z, \bar{z})+\bar{\Delta}(\bar{\partial}\bar{\epsilon})\phi(z, \bar{z}),
\eea
where 
\bea
\bar{\partial}\equiv\frac{\partial}{\partial\bar{z}}.
\eea
The $\bar{\epsilon}$ is an anti-holomorphic transformation parameter, and $\bar{\Delta}$ is a conformal dimension of the anti-holomorphic part. 

\subsubsection{Conformal Algebra}  
\noindent 
The conformal transformations are given by the following generators: 
\begin{itemize} 
\item 
Translation \ $P_{\mu}=-i\partial_{\mu}$; 
\item 
Lorentz Rotation \ $M_{\mu\nu}=-i(x_{\mu}\partial_{\nu}-x_{\nu}\partial_{\mu})$; 
\item
Dilaton \ $D=-ix^{\mu}\partial_{\mu}$;  
\item 
Special Conformal Transformation \ $K_{\mu}=-i\big(2x_{\mu}(x^{\nu}\partial_{\nu})-x^2\partial_{\mu}$\big). 
\end{itemize} 
The conformal algebra is in the following: 
\bea
\lbrack M_{\mu\nu}, P_{\rho}\rbrack
&=&
-i(\eta_{\nu\rho}P_{\mu}-\eta_{\mu\rho}P_{\nu}); 
\nn\\ 
\lbrack M_{\mu\nu}, K_{\rho}\rbrack
&=& 
-i(\eta_{\nu\rho}K_{\mu}-\eta_{\mu\rho}K_{\nu}); 
\nn\\
\lbrack M_{\mu\nu}, M_{\rho\sigma}\rbrack 
&=& 
-i(\eta_{\nu\rho}M_{\mu\sigma}
-\eta_{\mu\rho}M_{\nu\sigma}
+\eta_{\nu\sigma}M_{\mu\rho} 
-\eta_{\mu\sigma}M_{\rho\nu}); 
\nn\\ 
\lbrack D, P_{\mu}\rbrack
&=& 
iP_{\mu}; 
\nn\\ 
\lbrack D, K_{\mu}\rbrack
&=& 
-iK_{\mu}; 
\nn\\ 
\lbrack K_{\mu}, P_{\nu}\rbrack 
&=& 
2i(\eta_{\mu\nu}D+M_{\mu\nu}), 
\eea 
where 
\bea
\eta_{\mu\nu}=\mathrm{diag}(-1, 1, \cdots, 1). 
\eea
The group associated with the above algebra is isomorphic to SO(2, $d$). 
Therefore, the number of generators are 
\bea
C^{d+2}_2=\frac{(d+2)(d+1)}{2}
\eea
in CFT$_d$. 
When $d=2$, the conformal symmetry group SO(2, 2) is isomorphic to the gauge symmetry group SL(2)$\times$SL(2) of the gauge formulation. 
Due to the non-breaking of gauge symmetry by quantum effects, the gauge formulation's boundary description is CFT$_2$.  

\subsection{AdS$_3$ Solution}
\noindent 
The AdS$_{d+1}$ solution can be immersed in a ($d+2$)-dimensional flat spacetime 
\bea
ds_{d+1}^2=-dX_1^2-dX_2^2+\sum_{j=3}^{d+2}dX_j^2, 
\eea
in which the embedding coordinates satisfies 
\bea
-X_1^2-X_2^2+\sum_{j=3}^{d+2}X_j^2=\frac{d(d-1)}{2\Lambda}.
\eea 
The AdS$_{d+1}$ solution can be parametrized in the global coordinate:
\bea
X_1&=&\sqrt{\frac{-d(d-1)}{2\Lambda}}\cosh(\rho)\cos(\tau); 
\nn\\ 
X_2&=&\sqrt{\frac{-d(d-1)}{2\Lambda}}\cosh(\rho)\sin(\tau); 
\nn\\
X_j&=&\sqrt{\frac{-d(d-1)}{2\Lambda}}\sinh(\rho)\hat{x}_j, 
\eea
where 
\bea
\sum_{j=3}^{d+2}\hat{x}_j^2=1, 
\eea
and Poincaré coordinate: 
\bea
X_1&=&-\frac{d(d-1)}{4r\Lambda}\Bigg\lbrack 1+\frac{4r^2\Lambda^2}{d^2(d-1)^2}\Bigg(-\frac{d(d-1)}{2\Lambda}+\bigg(\sum_{j=3}^{d+1}x_j^2\bigg)-t^2\Bigg)
\Bigg\rbrack; 
\nn\\ 
X_2&=&\sqrt{-\frac{2\Lambda}{d(d-1)}}rt; 
\nn\\
X_j&=&\sqrt{-\frac{2\Lambda}{d(d-1)}}rx_j;
\nn\\ 
X_{d+2}&=&-\frac{d(d-1)}{4r\Lambda}\Bigg\lbrack 1-\frac{4r^2\Lambda^2}{d^2(d-1)^2}\Bigg(-\frac{d(d-1)}{2\Lambda}-\bigg(\sum_{j=3}^{d+1}x_j^2\bigg)+t^2\Bigg)
\Bigg\rbrack. 
\nn\\
\eea 
In the global coordinate, the range of parameters is: 
\bea
-\infty<\tau<\infty; \ \rho> 0. 
\eea
The range of parameters in the Poincaré coordinate is:
\bea
-\infty<t<\infty; \ r> 0. 
\eea 
We first introduce the properties of the AdS$_3$ metric in the Poincaré and global coordinates. 
We then demonstrate the AdS$_3$ metric in the global coordinate from the gauge formulation. 

\subsubsection{Poincaré Coordinate} 
\noindent 
We introduce the parameter 
\bea
z\equiv-\frac{1}{\Lambda r}, 
\eea
and then the AdS$_3$ metric becomes 
\bea
ds_{3P}^2=-\frac{1}{\Lambda z^2}(dz^2+dx_3^2-dt^2).  
\eea 
Due to $z>0$, the Poincaré coordinate only covers the upper half region. 
The asymptotic boundary of AdS$_3$ is at $z\rightarrow 0$. 

\subsubsection{Global Coordinate} 
\noindent 
We introduce the parameters:
\bea
r\equiv\sqrt{-\frac{1}{\Lambda}}\sinh(\rho); \ t\equiv \sqrt{-\frac{1}{\Lambda}}\tau, 
\eea
and then the AdS$_3$ metric becomes 
\bea
ds_{3G}^2=-(1-\Lambda r^2)dt^2
+\frac{1}{1-\Lambda r^2}dr^2
+r^2d\theta^2, 
\eea
where $0<\theta\le 2\pi$. 
This coordinate covers the entire AdS$_3$ spacetime, and is referred to as the global coordinate. 
The AdS$_3$ boundary is located at $r\rightarrow\infty$ and takes the form of a cylinder. 
The cylinder manifold is isomorphic to a sphere manifold by removing the top and bottom points. 
After doing the Wick rotation with the Euclidean time 
\bea
\psi\equiv it
\eea
and the identification 
\bea
z_t\equiv \theta+i\psi\sim z_t+2\pi\tau, 
\eea 
where $\tau$ is a complex structure, the boundary becomes a torus manifold. 

\subsubsection{Gauge Formulation} 
\noindent 
The equations of motion for $A_{\mu}$ and $\bar{A}_{\mu}$ in the gauge formulation ($S_{3DG}$) are: 
\bea
0&=&F_{\mu\nu}\equiv \partial_{\mu}A_{\nu}-\partial_{\nu}A_{\mu}+\lbrack A_{\mu}, A_{\nu}\rbrack; 
\nn\\ 
0&=&\bar{F}_{\mu\nu}\equiv \partial_{\mu}\bar{A}_{\nu}-\partial_{\nu}\bar{A}_{\mu}+\lbrack \bar{A}_{\mu}, \bar{A}_{\nu}\rbrack. 
\eea 
We can write the solutions in terms of the SL(2) transformations,  $g$ and $\bar{g}$: 
\bea
A=g^{-1}ag+g^{-1}dg; \ \bar{A}=\bar{g}^{-1}a\bar{g}+\bar{g}^{-1}d\bar{g}.  
\eea 
The field strength $\tilde{F}_{\mu\nu}$ associated to $a$ is zero: 
\bea
\tilde{F}_{\mu\nu}\equiv\partial_{\mu}a_{\nu}-\partial_{\nu}a_{\mu}+\lbrack a_{\mu}, a_{\nu}\rbrack=0. 
\eea
We choose the solution of the gauge fields with $F_{\mu\nu}=\bar{F}_{\mu\nu}=0$:  
\bea
lA&=&\sqrt{-\Lambda r^2+1}J_0dx^++\sqrt{-\Lambda}rJ_1dx^++\frac{dr}{\sqrt{-\Lambda r^2+1}}J_2; 
\nn\\ 
l\bar{A}&=&-\sqrt{-\Lambda r^2+1}J_0dx^-+\sqrt{-\Lambda}rJ_1dx^--\frac{dr}{\sqrt{-\Lambda r^2+1}}J_2, 
\eea
where $x^{\pm}\equiv t\pm\theta$, for obtaining the AdS$_3$ solution in the global coordinate. 
The vielbein can be determined by $A$ and $\bar{A}$ or Eq. \eqref{GVS}: 
\bea
e=\frac{l}{2}(A-\bar{A})\equiv e_+dx^++e_-dx^-+e_rdx^r. 
\eea
We can obtain the AdS$_3$ metric in the global coordinate by substituting the components of the vielbein to Eq. \eqref{MV}.

\subsection{Boundary Term in Gauge Formulation} 
\noindent 
To derive a non-trivial boundary description, we introduce a boundary term (at $r\rightarrow\infty$) to the gauge formulation. 
The Lagrangian description is 
\bea
&&S_{\mathrm{G}}
\nn\\
&=&\frac{k}{2\pi}\int d^3x\ \mathrm{Tr}\bigg(A_tF_{r\theta}-\frac{1}{2}\big(A_r\partial_tA_{\theta}-A_{\theta}\partial_tA_r\big)\bigg)
\nn\\
&&-\frac{k}{2\pi}\int d^3x\ \mathrm{Tr}\bigg(\bar{A}_t\bar{F}_{r\theta}-\frac{1}{2}\big(\bar{A}_r\partial_t\bar{A}_{\theta}-\bar{A}_{\theta}\partial_t\bar{A}_r\big)\bigg)
\nn\\
&&-\frac{k}{4\pi}\int dtd\theta\ \mathrm{Tr}\bigg(\frac{E_t^+}{E_{\theta}^+}A_{\theta}^2\bigg)
\nn\\
&&+\frac{k}{4\pi}\int dtd\theta\ \mathrm{Tr}\bigg(\frac{E_t^-}{E_{\theta}^-}\bar{A}_{\theta}^2\bigg)  
\eea
when the boundary zweibein $E$ is: 
\bea
E_t^+=-E_t^-=E_{\theta}^+=E_{\theta}^-=1.  
\eea 
The gauge fields satisfy the boundary conditions: 
\bea
(E_{\theta}^+A_t-E_t^+A_{\theta})|_{r\rightarrow\infty}=0; \qquad (E_{\theta}^-\bar{A}_t-E_t^-\bar{A}_{\theta})|_{r\rightarrow\infty}=0. 
\eea
The boundary metric can be rewritten in terms of $E$,  
\bea
g_{\tilde{\mu}\tilde{\nu}}\equiv\frac{1}{2}(E^+_{\tilde{\mu}}E^-_{\tilde{\nu}}+E^-_{\tilde{\mu}}E^+_{\tilde{\nu}}).  
\eea
The indices of boundary spacetimes are labeled as $\tilde{\mu}=t, \theta$. 
When utilizing the Euclidean signature, the boundary condition corresponds to the torus manifold. 
The boundary condition of other manifolds can be derived from the Weyl transformation \cite{Huang:2023aqz}. 

\subsection{2D Schwarzian Theory} 
\noindent 
We derive the 2D Schwarzian theory from the AdS$_3$ Einstein gravity in this section. 
The asymptotic AdS$_3$ boundary condition provides the boundary constraint \cite{Cotler:2018zff}. 
We then use the boundary description of Einstein gravity to show the 2D Schwarzian theory and discuss the modular symmetry. 
We demonstrate the dual of the 2D Schwarzian theory using chiral scalar fields.

\subsubsection{Boundary Constraint} 
\noindent 
The asymptotic behavior of the gauge fields is that: 
\bea
A|_{r\rightarrow\infty}=
\begin{pmatrix}
\frac{dr}{2r}& 0
\\
rE^+& -\frac{dr}{2r}
\end{pmatrix}; \ 
\bar{A}|_{r\rightarrow\infty}=
\begin{pmatrix}
-\frac{dr}{2r}&-rE^-
\\
0& \frac{dr}{2r}
\end{pmatrix}.
\eea 
We first parametrize the SL(2) transformations: 
\bea
g_{\mathrm{SL(2)}}&=&
\begin{pmatrix}
1& 0
\\
F& 1
\end{pmatrix}
\begin{pmatrix}
\lambda & 0
\\
0& \frac{1}{\lambda}
\end{pmatrix}
\begin{pmatrix}
1 &\Psi
\\
0& 1
\end{pmatrix}; 
\nn\\
\bar{g}_{\mathrm{SL(2)}}&=&
\begin{pmatrix}
1& -\bar{F}
\\
0& 1
\end{pmatrix}
\begin{pmatrix}
\frac{1}{\bar{\lambda}} & 0
\\
0& \bar{\lambda}
\end{pmatrix}
\begin{pmatrix}
1 &0
\\
-\bar{\Psi}& 1
\end{pmatrix}.
\eea 
By identifying the SL(2) transformations with the gauge fields: 
\bea
g^{-1}_{\mathrm{SL(2)}}\partial_{\theta}g_{\mathrm{SL(2)}}|_{r\rightarrow\infty}
=A_{\theta}|_{r\rightarrow\infty}, \qquad 
\bar{g}^{-1}_{\mathrm{SL(2)}}\partial_{\theta}\bar{g}_{\mathrm{SL(2)}}|_{r\rightarrow\infty}
=\bar{A}_{\theta}|_{r\rightarrow\infty}, 
\eea
we obtain the boundary constraint \cite{Cotler:2018zff}: 
\bea
\lambda=\sqrt{\frac{r E_{\theta}^+}{\partial_{\theta}F}}; \ 
\Psi=-\frac{1}{2rE_{\theta}^+}\frac{\partial_{\theta}^2F}{\partial_{\theta}F}, \qquad 
\bar{\lambda}=\sqrt{\frac{r E_{\theta}^-}{\partial_{\theta}\bar{F}}}; \  
\bar{\Psi}=-\frac{1}{2rE_{\theta}^-}\frac{\partial_{\theta}^2\bar{F}}{\partial_{\theta}\bar{F}}. 
\label{BC}
\eea

\subsubsection{SL(2) Measure}
\noindent
The boundary constraint \eqref{BC} has the symmetry generated by the composition of two SL(2) transformations:
\bea
&&
g_{\mathrm{SL(2)}}\rightarrow h_{\mathrm{SL(2)}}(t)g_{\mathrm{SL(2)}}, 
\nn\\
&& 
h_{\mathrm{SL(2)}}\equiv
\begin{pmatrix}
a_1(t)& a_2(t)
\\
a_3(t)&a_4(t)
\end{pmatrix}, \ a_1(t)a_4(t)-a_2(t)a_3(t)=1.
\eea
The new transformation can be written explicitly
\bea
h_{\mathrm{SL(2)}}g_{\mathrm{SL(2)}}=
\begin{pmatrix}
a_1\lambda+a_2F\lambda& a_1\lambda\Psi+a_2\bigg(F\lambda\Psi+\frac{1}{\lambda}\bigg)
\\
a_3\lambda+a_4F\lambda& a_3\lambda\Psi+a_4\bigg(F\lambda\Psi+\frac{1}{\lambda}\bigg)
\end{pmatrix}.
\eea
Hence we obtain the transformation:
\bea
\lambda&\rightarrow& a_1\lambda+a_2F\lambda, \qquad \lambda\Psi\rightarrow a_1\lambda\Psi+a_2\bigg(F\lambda\Psi+\frac{1}{\lambda}\bigg), 
\nn\\
 F\lambda&\rightarrow&a_3\lambda+a_4F\lambda, \qquad F\lambda\Psi+\frac{1}{\lambda}\rightarrow a_3\lambda\Psi+a_4\bigg(F\lambda\Psi+\frac{1}{\lambda}\bigg).
\eea
This transformation implies that the transformed fields are given by:
\bea
\tilde{\lambda}&=&a_1\lambda+a_2F\lambda, 
\nn\\
\tilde{\Psi}&=&\Psi+\frac{a_2}{(a_1+a_2F)\lambda^2}, 
\nn\\
\tilde{F}&=&\frac{a_4F+a_3}{a_2F+a_1}, 
\eea
We compute the SL(2) measure from the followings:
\bea
\frac{\partial g_{\mathrm{SL(2)}}}{\partial \lambda}&=&
\begin{pmatrix}
1& \Psi
\\
F& F\Psi-\frac{1}{\lambda^2}
\end{pmatrix}, 
\nn\\
\frac{\partial g_{\mathrm{SL(2)}}}{\partial F}&=&
\begin{pmatrix}
0& 0
\\
\lambda& \lambda\Psi
\end{pmatrix}, 
\nn\\
\frac{\partial g_{\mathrm{SL(2)}}}{\partial \Psi}&=&
\begin{pmatrix}
0& \lambda
\\
0& F\lambda
\end{pmatrix},
\eea
\bea
g^{-1}_{\mathrm{SL(2)}}\frac{\partial g_{\mathrm{SL(2)}}}{\partial \lambda}&=&
\begin{pmatrix}
\frac{1}{\lambda}+\Psi\lambda F& -\Psi\lambda
\\
-\lambda F &\lambda
\end{pmatrix}
\begin{pmatrix}
1& \Psi
\\
F& F\Psi-\frac{1}{\lambda^2}
\end{pmatrix}
=\begin{pmatrix}
\frac{1}{\lambda}&2\frac{\Psi}{\lambda}
\\
0& -\frac{1}{\lambda}
\end{pmatrix},
\nn\\
g^{-1}_{\mathrm{SL(2)}}\frac{\partial g_{\mathrm{SL(2)}}}{\partial F}&=&
\begin{pmatrix}
\frac{1}{\lambda}+\Psi\lambda F& -\Psi\lambda
\\
-\lambda F &\lambda
\end{pmatrix}
\begin{pmatrix}
0& 0
\\
\lambda& \lambda\Psi
\end{pmatrix}
=\begin{pmatrix}
-\lambda^2\Psi&-\lambda^2\Psi^2
\\
\lambda^2& \lambda^2\Psi
\end{pmatrix},
\nn\\
g^{-1}_{\mathrm{SL(2)}}\frac{\partial g_{\mathrm{SL(2)}}}{\partial \Psi}&=&
\begin{pmatrix}
\frac{1}{\lambda}+\Psi\lambda F& -\Psi\lambda
\\
-\lambda F &\lambda
\end{pmatrix}
\begin{pmatrix}
0& \lambda
\\
0& F\lambda
\end{pmatrix}
=\begin{pmatrix}
0&1
\\
0& 0
\end{pmatrix},
\eea
\bea
G_{\lambda\lambda}&\sim& \mathrm{Tr}\bigg(g^{-1}_{\mathrm{SL(2)}}\frac{\partial g_{\mathrm{SL(2)}}}{\partial \lambda}g^{-1}_{\mathrm{SL(2)}}\frac{\partial g_{\mathrm{SL(2)}}}{\partial \lambda}\bigg)=\frac{2}{\lambda^2},
\nn\\
G_{\lambda F}&=&G_{F\lambda}\sim\mathrm{Tr}\bigg(g^{-1}_{\mathrm{SL(2)}}\frac{\partial g_{\mathrm{SL(2)}}}{\partial \lambda}g^{-1}_{\mathrm{SL(2)}}\frac{\partial g_{\mathrm{SL(2)}}}{\partial F}\bigg)=0,
\nn\\
G_{\lambda\Psi}&=&G_{\Psi \lambda}\sim\mathrm{Tr}\bigg(g^{-1}_{\mathrm{SL(2)}}\frac{\partial g_{\mathrm{SL(2)}}}{\partial \lambda}g^{-1}_{\mathrm{SL(2)}}\frac{\partial g_{\mathrm{SL(2)}}}{\partial \Psi}\bigg)=0,
\nn\\
G_{FF}&\sim&\mathrm{Tr}\bigg(g^{-1}_{\mathrm{SL(2)}}\frac{\partial g_{\mathrm{SL(2)}}}{\partial F}g^{-1}_{\mathrm{SL(2)}}\frac{\partial g_{\mathrm{SL(2)}}}{\partial F}\bigg)=0,
\nn\\
G_{F\Psi}&=&G_{\Psi F}\sim\mathrm{Tr}\bigg(g^{-1}_{\mathrm{SL(2)}}\frac{\partial g_{\mathrm{SL(2)}}}{\partial F}g^{-1}_{\mathrm{SL(2)}}\frac{\partial g_{\mathrm{SL(2)}}}{\partial \Psi}\bigg)=\lambda^2,
\nn\\
G_{\Psi\Psi}&\sim&\mathrm{Tr}\bigg(g^{-1}_{\mathrm{SL(2)}}\frac{\partial g_{\mathrm{SL(2)}}}{\partial \Psi}g^{-1}_{\mathrm{SL(2)}}\frac{\partial g_{\mathrm{SL(2)}}}{\partial \Psi}\bigg)=0.
\eea
Hence the determinant of the matrix $G_{\mu\nu}$ gives  $-2\lambda^2$ term. 
The SL(2) measure is given by:
\bea
\int D g_{\mu\nu}\sim\int d\lambda\wedge dF\wedge d\Psi \sqrt{-\det G_{\mu\nu}} \sim \int d\lambda\wedge dF\wedge d\Psi\ \lambda.
\eea
Including the boundary constraint into the measure, we obtain the following result
\bea
\int d\lambda\wedge dF\wedge d\Psi\ \lambda \delta\bigg(\lambda^2 \big(\partial_{\theta}F\big)-rE_{\theta}^+\bigg)
\delta\bigg(\frac{1}{2rE_{\theta}^+}\frac{\partial_{\theta}^2F}{\partial_{\theta}F}+\Psi \bigg)\sim\int dF\ \frac{1}{\partial_{\theta} F}. 
\nn\\
\eea
We can also show that the measure is invariant under the SL(2) transformation 
\bea
\frac{dF}{\partial_{\theta} F}=\frac{d\tilde{F}}{\partial_{\theta} \tilde{F}}, 
\eea
where $\tilde{F}$ is the field $F$ after the transformation. 
This result also implies that the boundary constraint \eqref{BC} is invariant under the SL(2) transformation. 
Another SL(2) transformation $\bar{g}$ has a similar result. 

\subsubsection{2D Schwarzian Theory}
\noindent
Because the $A_t$ and $\bar{A}_t$ only has a linear coupling term in $S_G$,  we first integrate out the $A_t$, which is equivalent to using \cite{Cotler:2018zff}:
\bea
&&
F_{r\theta}=\bar{F}_{r\theta}=0; 
\nn\\
&&
g^{-1}_{\mathrm{SL(2)}}\partial_{\theta}g_{\mathrm{SL(2)}}=A_{\theta}, \ g^{-1}_{\mathrm{SL(2)}}\partial_{r}g_{\mathrm{SL(2)}}=A_{r},
\nn\\ 
&&
\bar{g}^{-1}_{\mathrm{SL(2)}}\partial_{\theta}\bar{g}_{\mathrm{SL(2)}}=\bar{A}_{\theta}, \ \bar{g}^{-1}_{\mathrm{SL(2)}}\partial_{r}\bar{g}_{\mathrm{SL(2)}}=\bar{A}_{r}; 
\eea 
\bea
&&S_{\mathrm{G}1}
\nn\\
&=&-\frac{k}{4\pi}\int d^3x\ \epsilon^{tr\theta}\mathrm{Tr}\bigg(-g_{\mathrm{SL(2)}}^{-1}\big(\partial_rg_{\mathrm{SL(2)}}\big)g_{\mathrm{SL(2)}}^{-1}\big(\partial_tg_{\mathrm{SL(2)}}\big)g_{\mathrm{SL(2)}}^{-1}\partial_{\theta}g_{\mathrm{SL(2)}}
\nn\\
&&
+g_{\mathrm{SL(2)}}^{-1}\big(\partial_rg_{\mathrm{SL(2)}}\big)g_{\mathrm{SL(2)}}^{-1}\big(\partial_{\theta}g_{\mathrm{SL(2)}}\big)g_{\mathrm{SL(2)}}^{-1}\big(\partial_tg_{\mathrm{SL(2)}}\big)
\bigg)
\nn\\
&&+\frac{k}{4\pi}\int d^3x\ \epsilon^{tr\theta}\mathrm{Tr}\bigg(-\bar{g}_{\mathrm{SL(2)}}^{-1}\big(\partial_r\bar{g}_{\mathrm{SL(2)}}\big)\bar{g}_{\mathrm{SL(2)}}^{-1}\big(\partial_t\bar{g}_{\mathrm{SL(2)}}\big)\bar{g}_{\mathrm{SL(2)}}^{-1}
\partial_{\theta}\bar{g}_{\mathrm{SL(2)}}
\nn\\
&&
+\bar{g}_{\mathrm{SL(2)}}^{-1}\big(\partial_r\bar{g}_{\mathrm{SL(2)}}\big)\bar{g}_{\mathrm{SL(2)}}^{-1}\big(\partial_{\theta}\bar{g}_{\mathrm{SL(2)}}\big)\bar{g}_{\mathrm{SL(2)}}^{-1}\big(\partial_t\bar{g}_{\mathrm{SL(2)}}\big)
\bigg)
\nn\\
&&+\frac{k}{2\pi}\int dt d\theta\ \mathrm{Tr}\bigg(g_{\mathrm{SL(2)}}^{-1}\big(\partial_{\theta}g_{\mathrm{SL(2)}}\big)g_{\mathrm{SL(2)}}^{-1}\big(D_-g_{\mathrm{SL(2)}}\big)\bigg)
\nn\\
&&-\frac{k}{2\pi}\int dt d\theta\ \mathrm{Tr}\bigg(\bar{g}_{\mathrm{SL(2)}}^{-1}\big(\partial_{\theta}\bar{g}_{\mathrm{SL(2)}}\big)\bar{g}_{\mathrm{SL(2)}}^{-1}\big(D_+\bar{g}_{\mathrm{SL(2)}}\big)\bigg),
\eea
in which we define that:
\bea
D_+=\frac{1}{2} \partial_t+\frac{1}{2}\frac{E_{t}^-}{E_{\theta}^-}\partial_{\theta}, \qquad D_-=\frac{1}{2}\partial_t+\frac{1}{2}\frac{E_{t}^+}{E_{\theta}^+}\partial_{\theta}.
\eea
\\

\noindent 
We can do the further computation to rewrite the action in terms of the fields:
\bea
&&-\frac{k}{4\pi}\int d^3x\ \epsilon^{tr\theta}\mathrm{Tr}\bigg(-g_{\mathrm{SL(2)}}^{-1}\big(\partial_rg_{\mathrm{SL(2)}}\big)g_{\mathrm{SL(2)}}^{-1}\big(\partial_tg_{\mathrm{SL(2)}}\big)g_{\mathrm{SL(2)}}^{-1}\partial_{\theta}g_{\mathrm{SL(2)}}
\nn\\
&&
+g_{\mathrm{SL(2)}}^{-1}\big(\partial_rg_{\mathrm{SL(2)}}\big)g_{\mathrm{SL(2)}}^{-1}\big(\partial_{\theta}g_{\mathrm{SL(2)}}\big)g_{\mathrm{SL(2)}}^{-1}\big(\partial_tg_{\mathrm{SL(2)}}\big)
\bigg)
\nn\\
&=&\frac{k}{4\pi}\int dtd\theta\ \lambda^2\bigg(\partial_{\theta} F\partial_t\Psi-\partial_t F\partial_{\theta}\Psi
\bigg);
\eea
\bea
&&\frac{k}{2\pi}\int dt d\theta\ \mathrm{Tr}\bigg(g^{-1}\big(\partial_{\theta}g\big)g^{-1}\big(D_-g\big)\bigg)
\nn\\
&=&\frac{k}{2\pi}\int dtd\theta\ \bigg(\frac{2}{\lambda^2}(\partial_{\theta}\lambda)(D_-\lambda)+\lambda^2\big((D_-F)(\partial_{\theta}\Psi)+(\partial_{\theta}F)(D_-\Psi)\big)
\bigg);
\eea
\bea
&&-\frac{k}{4\pi}\int d^3x\ \epsilon^{tr\theta}\mathrm{Tr}\bigg(-g_{\mathrm{SL(2)}}^{-1}\big(\partial_rg_{\mathrm{SL(2)}}\big)g_{\mathrm{SL(2)}}^{-1}\big(\partial_tg_{\mathrm{SL(2)}}\big)g_{\mathrm{SL(2)}}^{-1}\partial_{\theta}g_{\mathrm{SL(2)}}
\nn\\
&&
+g_{\mathrm{SL(2)}}^{-1}\big(\partial_rg_{\mathrm{SL(2)}}\big)g_{\mathrm{SL(2)}}^{-1}\big(\partial_{\theta}g_{\mathrm{SL(2)}}\big)g_{\mathrm{SL(2)}}^{-1}\big(\partial_tg_{\mathrm{SL(2)}}\big)
\bigg)
\nn\\
&&+
\frac{k}{2\pi}\int dt d\theta\ \mathrm{Tr}\bigg(g_{\mathrm{SL(2)}}^{-1}\big(\partial_{\theta}g_{\mathrm{SL(2)}}\big)g_{\mathrm{SL(2)}}^{-1}\big(D_-g_{\mathrm{SL(2)}}\big)\bigg)
\nn\\
&=&\frac{k}{2\pi}\int dtd\theta\ 
\bigg(\frac{2}{\lambda^2}(\partial_{\theta}\lambda)(D_-\lambda)+2\lambda^2(\partial_{\theta}F)(D_-\Psi)\bigg)
\nn\\
&=&\frac{k}{\pi}\int dtd\theta\ \bigg(\frac{(\partial_{\theta}\lambda)(D_-\lambda)}{\lambda^2}+\lambda^2(\partial_{\theta}F)(D_-\Psi)\bigg).
\eea 
We can also get a similar result from the $\bar{g}$ \cite{Cotler:2018zff}. 
Therefore, we acquire the boundary description \cite{Cotler:2018zff} 
\bea
&&S_{\mathrm{G}1}
\nn\\
&=&\frac{k}{\pi}\int dtd\theta\ \bigg(\frac{(\partial_{\theta}\lambda)(D_-\lambda)}{\lambda^2}+\lambda^2(\partial_{\theta}F)(D_-\Psi)\bigg)
\nn\\
&&
-\frac{k}{\pi}\int dtd\theta\ \bigg(\frac{(\partial_{\theta}\bar{\lambda})(D_+\bar{\lambda})}{\bar{\lambda}^2}+\bar{\lambda}^2(\partial_{\theta}\bar{F})(D_+\bar{\Psi})\bigg).
\label{WZ}
\eea
After inserting the boundary constraint \eqref{BC}, we can rewrite the boundary Lagrangian in terms of $F$ and $\bar{F}$ \cite{Cotler:2018zff},
 \bea
&&S_{\mathrm{G}1}
\nn\\
&=&\frac{k}{2\pi}\int dtd\theta\ \bigg(\frac{3}{2}\frac{(D_-\partial_{\theta}{\cal F})(\partial_{\theta}^2{\cal F})}{(\partial_{\theta}{\cal F})^2}-\frac{D_-\partial_{\theta}^2{\cal F}}{\partial_{\theta}{\cal F}}
\bigg)
\nn\\
&&-\frac{k}{2\pi}\int dtd\theta\ \bigg(\frac{3}{2}\frac{(D_+\partial_{\theta}\bar{{\cal F}})(\partial_{\theta}^2\bar{{\cal F}})}{(\partial_{\theta}\bar{{\cal F}})^2}-\frac{D_+\partial_{\theta}^2\bar{{\cal F}}}{\partial_{\theta}\bar{{\cal F}}}
\bigg),  
\label{2DS}
\eea 
where the new variables ${\cal F}$ and $\bar{{\cal F}}$ are given by \cite{Cotler:2018zff}:  
\bea
{\cal F}\equiv\frac{F}{E^+_{\theta}}; \qquad
\bar{{\cal F}}\equiv\frac{\bar{F}}{E_{\theta}^-}.
\eea
The theory is called 2D Schwarzian theory after the Wick rotation (using the Euclidean time) and the identification ($z_t\sim z_t+2\pi\tau$). 
The constant solutions of $E^{\pm}$ do not break the form of the 2D Schwarzian theory. 
However, the general Weyl transformation cannot guarantee the same form of the boundary Lagrangian. 

\subsubsection{Modular Symmetry} 
\noindent
Because we have the boundary condition: 
\bea
\lambda^2\partial_{\theta}F=E_{\theta}^+r; \qquad \bar{\lambda}^2\partial_{\theta}\bar{F}2E_{\theta}^-r,
\eea
the terms that follow are considered as total derivatives in Eq. \eqref{WZ} \cite{Cotler:2018zff}: 
\bea
\frac{k}{\pi}\int dtd\theta\ \bigg(\lambda^2(\partial_{\theta}F)(D_-\Psi)\bigg), \qquad
-\frac{k}{\pi}\int dtd\theta\ \bigg(\bar{\lambda}^2(\partial_{\theta}\bar{F})(D_+\bar{\Psi})\bigg).
\eea
Therefore, the boundary theory becomes \cite{Cotler:2018zff}
\bea
S_{\mathrm{G}1}
=\frac{k}{\pi}\int dtd\theta\ \bigg(\frac{(\partial_{\theta}\lambda)(D_-\lambda)}{\lambda^2}
-\frac{(\partial_{\theta}\bar{\lambda})(D_+\bar{\lambda})}{\bar{\lambda}^2}\bigg).
\eea 
Because $E_{\theta}^{\pm}$ is a constant, we can simplify the expression of $\lambda$ and $\bar{\lambda}$ \cite{Cotler:2018zff}:
\bea
\lambda=\sqrt{\frac{r}{\partial_{\theta}{\cal F}}}; \qquad
\bar{\lambda}=\sqrt{\frac{r}{\partial_{\theta}\bar{{\cal F}}}}.
\eea
The derivative acting on the $\lambda$ provides:
\bea
\partial_{\theta}\lambda=-\frac{\sqrt{r}}{2}\frac{\partial_{\theta}^2{\cal F}}{(\partial_{\theta}{\cal F})^{\frac{3}{2}}}, \qquad
D_-\lambda=-\frac{\sqrt{r}}{2}\frac{D_-\partial_{\theta}{\cal F}}{(\partial_{\theta}{\cal F})^{\frac{3}{2}}}.
\eea
Therefore, we obtain:
\bea
(\partial_{\theta}\lambda)(D_-\lambda)=\frac{r}{4}\frac{\partial_{\theta}^2{\cal F}D_-\partial_{\theta}{\cal F}}{(\partial_{\theta}{\cal F})^3}, \qquad
\frac{(\partial_{\theta}\lambda)(D_-\lambda)}{\lambda^2}=\frac{1}{4}\frac{(\partial_{\theta}^2{\cal F})(D_-\partial_{\theta}{\cal{F}})}{(\partial_{\theta}{\cal F})^2}.
\eea
Therefore, we acquire an alternative way to describe the boundary \cite{Cotler:2018zff}
\bea
S_{\mathrm{G}1}
=\frac{k}{4\pi}\int dtd\theta\ \bigg(\frac{(\partial_{\theta}^2{\cal F})(D_-\partial_{\theta}{\cal{F}})}{(\partial_{\theta}{\cal F})^2}
-\frac{(\partial_{\theta}^2\bar{{\cal F}})(D_+\partial_{\theta}\bar{{\cal{F}}})}{(\partial_{\theta}\bar{{\cal F}})^2}\bigg).
\eea
\\

\noindent
Finally, we choose the field redefinition \cite{Cotler:2018zff}:
\bea
{\cal F}\equiv \tan\bigg(\frac{\phi}{2}\bigg); \qquad \bar{{\cal F}}\equiv\tan\bigg(\frac{\bar{\phi}}{2}\bigg)
\eea
to rewrite the following term \cite{Cotler:2018zff}
\bea
\frac{k}{4\pi}\int dtd\theta\ \frac{(\partial_{\theta}^2{\cal F})(D_-\partial_{\theta}{\cal{F}})}{(\partial_{\theta}{\cal F})^2}
=\frac{k}{4\pi}\int dtd\theta\ \bigg\lbrack\frac{(\partial_{\theta}^2\phi)(D_-\partial_{\theta}\phi)}{(\partial_{\theta}\phi)^2}
-(\partial_{\theta}\phi)(D_-\phi)\bigg\rbrack
\nn\\
\eea
by using:
\bea
&&\partial_{\theta}{\cal F}
\nn\\
&=&\frac{1}{2}\sec^2\bigg(\frac{\phi}{2}\bigg)(\partial_{\theta}\phi),
\nn\\
&&\partial_{\theta}^2{\cal F}
\nn\\
&=&\frac{1}{2}\sec^3\bigg(\frac{\phi}{2}\bigg)\sin\bigg(\frac{\phi}{2}\bigg)(\partial_{\theta}\phi)^2
+\frac{1}{2}\sec^2\bigg(\frac{\phi}{2}\bigg)(\partial_{\theta}^2\phi), 
\nn\\
&&\partial_t\partial_{\theta}{\cal F}
\nn\\
&=&\frac{1}{2}\sec^3\bigg(\frac{\phi}{2}\bigg)\sin\bigg(\frac{\phi}{2}\bigg)(\partial_t\phi)(\partial_{\theta}\phi)
+\frac{1}{2}\sec^2\bigg(\frac{\phi}{2}\bigg)(\partial_t\partial_{\theta}\phi),
\nn\\
&&\partial_{\theta}D_-{\cal F}
\nn\\
&=&\frac{1}{2}\sec^3\bigg(\frac{\phi}{2}\bigg)\sin\bigg(\frac{\phi}{2}\bigg)(\partial_{\theta}\phi)(D_-\phi)
+\frac{1}{2}\sec^2\bigg(\frac{\phi}{2}\bigg)(D_-\partial_{\theta}\phi),
\nn\\
&&(\partial_{\theta}^2{\cal F})(D_-\partial_{\theta}{\cal F})
\nn\\
&=&\frac{1}{4}\sec^6\bigg(\frac{\phi}{2}\bigg)\sin^2\bigg(\frac{\phi}{2}\bigg)(\partial_{\theta}\phi)^3D_-\phi
\nn\\
&&+\frac{1}{4}\sec^5\bigg(\frac{\phi}{2}\bigg)\sin\bigg(\frac{\phi}{2}\bigg)(\partial_{\theta}\phi)^2(D_-\partial_{\theta}\phi)
\nn\\
&&+\frac{1}{4}\sec^5\bigg(\frac{\phi}{2}\bigg)\sin\bigg(\frac{\phi}{2}\bigg)(\partial_{\theta}\phi)(\partial_{\theta}^2\phi)(D_-\phi)
\nn\\
&&+\frac{1}{4}\sec^4\bigg(\frac{\phi}{2}\bigg)(D_-\partial_{\theta}\phi)(\partial_{\theta}^2\phi),
\nn\\
&&\frac{(\partial_{\theta}^2{\cal F})(D_-\partial_{\theta}{\cal F})}{(\partial_{\theta}{\cal F})^2}
\nn\\
&=&\sec^2\bigg(\frac{\phi}{2}\bigg)\sin^2\bigg(\frac{\phi}{2}\bigg)(\partial_{\theta}\phi)(D_-\phi)
\nn\\
&&+\sec\bigg(\frac{\phi}{2}\bigg)\sin\bigg(\frac{\phi}{2}\bigg)(D_-\partial_{\theta}\phi)
\nn\\
&&+\sec\bigg(\frac{\phi}{2}\bigg)\sin\bigg(\frac{\phi}{2}\bigg)\frac{\partial_{\theta}^2\phi D_-\phi}{\partial_{\theta}\phi}
\nn\\
&&+\frac{(\partial_{\theta}^2\phi)(D_-\partial_{\theta}\phi)}{(\partial_{\theta}\phi)^2},
\nn
\eea
\bea
&&\int dtd\theta\ \frac{(\partial_{\theta}^2{\cal F})(D_-\partial_{\theta}{\cal F})}{(\partial_{\theta}{\cal F})^2}
\nn\\
&=&\int dtd\theta\ \bigg\lbrack\tan^2\bigg(\frac{\phi}{2}\bigg)(\partial_{\theta}\phi)(D_-\phi)
-\frac{1}{2}\sec^2\bigg(\frac{\theta}{2}\bigg)(\partial_{\theta}\phi)(D_-\phi)
\nn\\
&&-\frac{1}{2}\sec^2\bigg(\frac{\phi}{2}\bigg)(D_-\phi)(\partial_{\theta}\phi)
+\tan\bigg(\frac{\phi}{2}\bigg)\frac{(\partial_{\theta}^2\phi)(D_-\phi)}{\partial_{\theta}\phi}
-\tan\bigg(\frac{\phi}{2}\bigg)(D_-\partial_{\theta}\phi)
\nn\\
&&
+\frac{(\partial_{\theta}^2\phi)(D_-\partial_{\theta}\phi)}{(\partial_{\theta}\phi)^2}
\bigg\rbrack
\nn\\
&=&\int dtd\theta\ \bigg\lbrack\frac{(\partial_{\theta}^2\phi)(D_-\partial_{\theta}\phi)}{(\partial_{\theta}\phi)^2}
-(\partial_{\theta}\phi)(D_-\phi)
\nn\\
&&+\tan\bigg(\frac{\phi}{2}\bigg)\frac{(\partial_{\theta}^2\phi)(D_-\phi)}{\partial_{\theta}\phi}
-\tan\bigg(\frac{\phi}{2}\bigg)(D_-\partial_{\theta}\phi)\bigg\rbrack
\nn\\
&=&\int dtd\theta\ \bigg\lbrack\frac{(\partial_{\theta}^2\phi)(D_-\partial_{\theta}\phi)}{(\partial_{\theta}\phi)^2}
-(\partial_{\theta}\phi)(D_-\phi)
\nn\\
&&+\frac{1}{2}\tan\bigg(\frac{\phi}{2}\bigg)\frac{(\partial_{\theta}^2\phi)(\partial_t\phi)}{\partial_{\theta}\phi}
-\frac{1}{2}\tan\bigg(\frac{\phi}{2}\bigg)(\partial_t\partial_{\theta}\phi)\bigg\rbrack
\nn\\
&=&\int dtd\theta\ \bigg\lbrack\frac{(\partial_{\theta}^2\phi)(D_-\partial_{\theta}\phi)}{(\partial_{\theta}\phi)^2}
-(\partial_{\theta}\phi)(D_-\phi)\bigg\rbrack,
\eea
in which we use the following result in the last equality:
\bea
&&\int dtd\theta\ \bigg\lbrack\tan\bigg(\frac{\phi}{2}\bigg)\frac{(\partial_{\theta}^2\phi)(\partial_t\phi)}{\partial_{\theta}\phi}
-\tan\bigg(\frac{\phi}{2}\bigg)(\partial_t\partial_{\theta}\phi)\bigg\rbrack
\nn\\
&=&\int dtd\theta\ \bigg\{-\ln(\partial_{\theta}\phi)\partial_{\theta}\bigg\lbrack\partial_t{\phi}\tan\bigg(\frac{\phi}{2}\bigg)\bigg\rbrack
+\ln(\partial_{\theta}\phi)\partial_t\bigg\lbrack(\partial_{\theta}\phi)\tan\bigg(\frac{\phi}{2}\bigg)\bigg\rbrack
\nn\\
&&+\partial_{\theta}\bigg\lbrack\ln(\partial_{\theta}\phi)(\partial_t\phi)\tan\bigg(\frac{\phi}{2}\bigg)\bigg\rbrack
-\partial_t\bigg\lbrack\ln(\partial_{\theta}\phi)(\partial_{\theta}\phi)\tan\bigg(\frac{\phi}{2}\bigg)\bigg\rbrack\bigg\}
\nn\\
&=&\int dtd\theta\ \bigg\{\partial_{\theta}\bigg\lbrack\ln(\partial_{\theta}\phi)(\partial_t\phi)\tan\bigg(\frac{\phi}{2}\bigg)\bigg\rbrack
-\partial_t\bigg\lbrack\ln(\partial_{\theta}\phi)(\partial_{\theta}\phi)\tan\bigg(\frac{\phi}{2}\bigg)\bigg\rbrack\bigg\}
\nn\\
&=&0.
\eea 
We can obtain a similar result from $\bar{F}$ \cite{Cotler:2018zff}. 
Therefore, the Lagrangian description becomes \cite{Cotler:2018zff}:
\bea
&&S_{\mathrm{G}1}
\nn\\
&=&\frac{k}{4\pi}\int dtd\theta\ \bigg(\frac{(\partial_{\theta}^2{\cal F})(D_-\partial_{\theta}{\cal{F}})}{(\partial_{\theta}{\cal F})^2}
-\frac{(\partial_{\theta}^2\bar{{\cal F}})(D_+\partial_{\theta}\bar{{\cal{F}}})}{(\partial_{\theta}\bar{{\cal F}})^2}\bigg)
\nn\\
&=&\frac{k}{4\pi}\int dtd\theta\ \bigg\lbrack\frac{(\partial_{\theta}^2\phi)(D_-\partial_{\theta}\phi)}{(\partial_{\theta}\phi)^2}
-(\partial_{\theta}\phi)(D_-\phi)\bigg\rbrack
\nn\\
&&
-\frac{k}{4\pi}\int dtd\theta\ \bigg\lbrack\frac{(\partial_{\theta}^2\bar{\phi})(D_+\partial_{\theta}\bar{\phi})}{(\partial_{\theta}\bar{\phi})^2}
-(\partial_{\theta}\bar{\phi})(D_+\bar{\phi})\bigg\rbrack.
\label{OE}
\eea 
The measure becomes \cite{Cotler:2018zff}
\bea
\int\frac{d\phi}{\partial_{\theta}\phi}\frac{d\bar{\phi}}{\partial_{\theta}\bar{\phi}}
\eea
 The path integral over $\phi$ and $\bar{\phi}$ that we identify \cite{Cotler:2018zff}: 
 \bea
 \tan\bigg(\frac{\phi}{2}\bigg)&\sim& \frac{a_4(\psi)\tan\big(\frac{\phi}{2}\big)+a_3(\psi)}{a_2(\psi)\tan\big(\frac{\phi}{2}\big)+a_1(\psi)}; 
 \nn\\ 
 \tan\bigg(\frac{\bar{\phi}}{2}\bigg)&\sim& \frac{\bar{a}_4(\psi)\tan\big(\frac{\bar{\phi}}{2}\big)+\bar{a}_3(\psi)}{\bar{a}_2(\psi)\tan\big(\frac{\bar{\phi}}{2}\big)+\bar{a}_1(\psi)}. 
 \eea
\\

\noindent 
 The on-shell-solution is unique up to the gauge redundancy \cite{Cotler:2018zff}
 \bea
 \phi_0=\theta-\frac{\mathrm{Re}(\tau)}{\mathrm{Im}(\tau)}\psi. 
 \eea
 We then obtain the boundary condition for the A-cycle of the torus \cite{Cotler:2018zff}: 
 \bea
 &&
 \phi(\psi, \theta+2\pi)=\phi(\psi, \theta)+2\pi;  
 \nn\\
 &&
 \phi\big(\psi+2\pi\mathrm{Im}(\tau), \theta+2\pi\mathrm{Re}(\tau)\big)=\phi(\psi, \theta).  
 \eea
 For the B-cycle, the bulk geometry is the Euclidean BTZ black hole. 
 The solution is also unique \cite{Cotler:2018zff}
 \bea
 \phi_0=\frac{\psi}{\mathrm{Im}(\tau)}. 
 \eea
 The boundary condition is \cite{Cotler:2018zff}:
\bea
&&
 \phi(\psi, \theta+2\pi)=\phi(\psi, \theta);
 \nn\\
 && 
 \phi\big(\psi+2\pi\mathrm{Im}(\tau), \theta+2\pi\mathrm{Re}(\tau)\big)=\phi(\psi, \theta)+2\pi .  
 \eea
 The path integral over $\phi$ and $\bar{\phi}$ that we identify \cite{Cotler:2018zff}: 
 \bea
 \tan\bigg(\frac{\phi}{2}\bigg)\sim \frac{\tilde{a}_4\tan\big(\frac{\phi}{2}\big)+\tilde{a}_3}{\tilde{a}_2\tan\big(\frac{\phi}{2}\big)+\tilde{a}_1}; \ 
 \tan\bigg(\frac{\bar{\phi}}{2}\bigg)\sim \frac{\bar{\tilde{a}}_4\tan\big(\frac{\bar{\phi}}{2}\big)+\bar{\tilde{a}}_3}{\bar{\tilde{a}}_2\tan\big(\frac{\bar{\phi}}{2}\big)+\bar{\tilde{a}}_1(\psi)}, 
 \eea
 where $\tilde{a}_1$, $\tilde{a}_2$, $\tilde{a}_3$, $\tilde{a}_4$, $\bar{\tilde{a}}_1$, $\bar{\tilde{a}}_2$, $\bar{\tilde{a}}_3$, $\bar{\tilde{a}}_4$ are functions of 
 \bea
 \theta-\frac{\mathrm{Re}(\tau)}{\mathrm{Im}(\tau)}\psi
 \eea
 satisfying \cite{Cotler:2018zff}
 \bea
 \tilde{a}_1\tilde{a}_4-\tilde{a}_2\tilde{a}_3=\bar{\tilde{a}}_1\bar{\tilde{a}}_4-\bar{\tilde{a}}_2\bar{\tilde{a}}_3=1.
 \eea 
 $\bar{\psi}$ has a similar result for the A-cycle and B-cycle \cite{Cotler:2018zff}.  
The modular transformation 
\bea
\tau\rightarrow-\frac{1}{\tau}, 
\eea 
swaps the A-cycle and the B-cycle. 
The A-cycle and B-cycle partition functions are not invariant under modular transformation, indicating a lack of modular symmetry in this theory \cite{Cotler:2018zff}. 
The loss of modular symmetry is due to the non-periodic boundary condition. 
The conventional CFT on the torus requires periodic boundary conditions for each direction. 
Hence the CFT$_2$ without the modular symmetry on the torus is the boundary theory of AdS$_3$ Einstein gravity theory. 

\subsubsection{Chiral Scalar Fields}
\noindent
 Now we show that 2D Schwarzian theory is dual to the following action \cite{Huang:2020tjl}
\bea
S_{2D1}=\frac{4k}{\pi}\int dt d\theta\ \bigg(\big(D_-\phi\big)\big(\partial_{\theta}\phi\big)+\Pi\big(\partial_{\theta}{\cal F}-e^{4\phi}\big)\bigg).
\eea
The measure of path integration is 
\bea
\int d\phi d{\cal F} d\Pi.
\eea 
If we first integrate out the $\Pi$ and then integrate out the $\phi$, equivalent to replacing $\phi$ by ${\cal F}$ with the following equality
\bea
\ln\partial_{\theta}{\cal F}=4\phi,
\eea
and then we obtain 
\bea
S_{2D2}=\frac{k}{4\pi}\int dtd\theta\  \frac{\partial_{\theta}^2{\cal F}}{(\partial_{\theta}{\cal F})^2}\big(D_-\partial_{\theta}{\cal F}\big).
\eea
The measure becomes 
\bea
\int \frac{d{\cal F}}{\partial_{\theta}{\cal F}}.
\eea
We can show that the dual theory is equivalent to the 2D Schwarzian theory up to a total derivative term (integration by part in $\theta$, and the total derivative term vanishes for the torus manifold) \cite{Huang:2020tjl}
\bea
\frac{k}{2\pi}\int dtd\theta\ \bigg(\frac{3}{2}\frac{(D_-\partial_{\theta}{\cal F})(\partial_{\theta}^2{\cal F})}{(\partial_{\theta}{\cal F})^2}-\frac{D_-\partial_{\theta}^2{\cal F}}{\partial_{\theta}{\cal F}}\bigg)
=\frac{k}{4\pi}\int dtd\theta\ \frac{\partial_{\theta}^2{\cal F}}{(\partial_{\theta}{\cal F})^2}(D_-\partial_{\theta}{\cal F}). 
\nn\\
\eea 
\\

\noindent
Now we integrate out the ${\cal F}$ in $S_{2D1}$, which is equivalent to introducing a constraint to the measure \cite{Huang:2020tjl}
\bea
\delta\big(\Pi-f(t)\big).
\eea 
We then can integrate out $\Pi$ and perform a field redefinition
\bea
\phi\rightarrow\phi-\frac{1}{4}\ln f
\eea
to obtain the following action \cite{Huang:2020tjl}
\bea
S_{2D3}=\frac{2k}{\pi}\int dtd\theta\ \bigg(\big(\partial_t\phi\big)\big(\partial_{\theta}\phi\big)
-\frac{E_t^+}{E_{\theta}^+}\big(\partial_{\theta}\phi\big)\big(\partial_{\theta}\phi\big)
-e^{4\phi}\bigg),
\eea
up to a total derivative term. 
The measure becomes 
\bea
\int d\phi.
\eea
We can obtain a similar dual for the $\bar{A}$ \cite{Huang:2020tjl}. 
Therefore, the dual of the 2D Schwarzian theory is represented by the chiral scalar fields below \cite{Huang:2020tjl}
\bea
&&
S_{CB}
\nn\\
&=&
\frac{4k}{\pi}\int dt d\theta\ \bigg(\big(D_-\phi\big)\big(\partial_{\theta}\phi\big)-e^{4\phi}\bigg)
-\frac{4k}{\pi}\int dt d\theta\ \bigg(\big(D_+\bar{\phi}\big)\big(\partial_{\theta}\bar{\phi}\big)+e^{4\bar{\phi}}\bigg).
\nn\\
\eea
The measure is 
\bea
\int d\phi d\bar{\phi}.
\eea 
We can observe the difference between the Liouville theory and 2D Schwarzian theory on the torus manifold \cite{Huang:2020tjl}. 

\subsection{Weyl Transformation and Liouville Theory}
\noindent 
Because the Lagrangian description of 2D Schwarzian theory loses the covariant form, we cannot use the conventional result of Weyl anomaly from CFT$_2$ \cite{Cotler:2018zff}. 
However, the direct calculation for the torus manifold shows the Liouville theory resulting from the Weyl transformation \cite{Huang:2023aqz}.  
The spin connection $\omega^a$ satisfies the torsionless condition (or equation of motion of $\omega^a$)
\bea
de_a+\epsilon_{abc}\omega^b\wedge e^c=0. 
\eea
We can solve the spin connection, and the asymptotic solution is: 
\bea
&&
e^0|_{r\rightarrow\infty}=rdt, \ 
e^1|_{r\rightarrow\infty}=rd\theta, \ 
e^2|_{r\rightarrow\infty}=0;  
\nn\\
&&
\omega^0|_{r\rightarrow\infty}=rd\theta,\ 
\omega^1|_{r\rightarrow\infty}=rdt, \ 
\omega^2|_{r\rightarrow\infty}=0.
\eea
\\

\noindent
A general Weyl transformation 
\bea
e^a\rightarrow\exp\big(\sigma(t, \theta)\big)e^a  
\eea
leads the asymptotic boundary condition to the gauge fields \cite{Huang:2023aqz}: 
\bea
A_{\theta}|_{r\rightarrow\infty}=
\begin{pmatrix}
-\frac{1}{2}\partial_t\sigma& 0
\\
re^{\sigma}E_{\theta}^+& \frac{1}{2}\partial_t\sigma 
\end{pmatrix}; \ 
\bar{A}_{\theta}|_{r\rightarrow\infty}=
\begin{pmatrix}
-\frac{1}{2}\partial_t\sigma& -re^{\sigma}E_{\theta}^-
\\
0& \frac{1}{2}\partial_t\sigma  
\end{pmatrix}.
\eea
The boundary constraints after the Weyl transformation become \cite{Huang:2023aqz}: 
\bea
\lambda&=&e^{\frac{\sigma}{2}}\lambda^+,
\nn\\ 
\Psi&=&-\frac{e^{-\sigma}}{2E_{\theta}^+r}\big(\partial_{\theta}(\ln{\cal F}^+)-\partial_{\theta}\sigma-\partial_t\sigma\big), 
\nn\\ 
{\cal F}&\equiv& e^{-\sigma}{\cal F}^+;  
\nn\\
\bar{\lambda}&=&e^{\frac{\sigma}{2}}\lambda^-, 
\nn\\ 
\bar{\Psi}&=&-\frac{e^{-\sigma}}{2E_{\theta}^-r}\big(\partial_{\theta}(\ln{\cal F}^-)-\partial_{\theta}\sigma+\partial_t\sigma\big), 
\nn\\ 
\bar{{\cal F}}&\equiv& e^{-\sigma}{\cal F}^-.
\eea 
where 
\bea
&&
\lambda^{+}\equiv\sqrt{\frac{rE_{\theta}^+}{\partial_{\theta}F}}, \ 
\lambda^{-}\equiv\sqrt{\frac{rE_{\theta}^-}{\partial_{\theta}\bar{F}}}; 
\nn\\
&& 
{\cal F}^+\equiv\frac{\partial_{\theta}F}{E_{\theta}^+}, \ 
{\cal F}^-\equiv\frac{\partial_{\theta}\bar{F}}{E_{\theta}^-}. 
\eea
The boundary constraints lead to the different boundary conditions \cite{Huang:2023aqz}:
\bea
&&
(E_{\theta}^+A_t-E_t^+A_{\theta}-E_{\theta}^+A_t^2J_2+E_t^+A_{\theta}^2J_2)|_{r\rightarrow\infty}=0; 
\nn\\
&& 
(E_{\theta}^-\bar{A}_t-E_t^-\bar{A}_{\theta}-E_{\theta}^-\bar{A}_t^2+E_t^-\bar{A}_{\theta}^2J_2)|_{r\rightarrow\infty}=0, 
\eea
where 
\bea
A_{\theta}^2=\bar{A}_{\theta}^2=-\partial_t\sigma; \ 
A_t^2=\bar{A}_t^2=-\partial_{\theta}\sigma.  
\eea 
The upper index of $A_{\theta}^2$ is the Lie algebra index. 
\\

\noindent
The variation of $A_{\theta}$ and $\bar{A}_{\theta}$ \cite{Huang:2023aqz}
\bea
&&
-\frac{k}{8\pi}\int dt d\theta\ \bigg(\frac{E_t^+}{E_{\theta}^+}\delta(A_{\theta}^2A_{\theta}^2)
-\delta(A_{t}^2A_{t}^2)\bigg)
\nn\\
&&
+\frac{k}{8\pi}\int dt d\theta\ \bigg(\frac{E_t^-}{E_{\theta}^-}\delta(\bar{A}_{\theta}^2\bar{A}_{\theta}^2)
-\delta(\bar{A}_{t}^2\bar{A}_{t}^2)\bigg) 
\nn\\
\eea 
shows the necessity of introducing the additional boundary term \cite{Huang:2023aqz}: 
\bea
&&
S_{\mathrm{B}1}
\nn\\
&=&\frac{k}{8\pi}\int dt d\theta\ \frac{E_t^+}{E_{\theta}^+}A_{\theta}^2A_{\theta}^2
-\frac{k}{8\pi}\int dtd\theta\ A_t^2A_t^2
\nn\\
&&
-\frac{k}{8\pi}\int dt d\theta\ \frac{E_t^-}{E_{\theta}^-}\bar{A}_{\theta}^2\bar{A}_{\theta}^2
+\frac{k}{8\pi}\int dtd\theta\ \bar{A}_t^2\bar{A}_t^2
\nn\\
&=&\frac{k}{8\pi}\int dt d\theta\ \frac{E_t^+}{E_{\theta}^+}A_{\theta}^2A_{\theta}^2
-\frac{k}{8\pi}\int dt d\theta\ \frac{E_t^-}{E_{\theta}^-}\bar{A}_{\theta}^2\bar{A}_{\theta}^2
\nn\\
&=&\frac{k}{4\pi}\int dtd\theta\ (\partial_t\sigma)(\partial_t\sigma)
. 
\eea 
Applying the Weyl transformation to the boundary action \eqref{WZ} generates the additional term \cite{Huang:2023aqz}: 
\bea
&&
\frac{(\partial_{\theta}\lambda)(D_-\lambda)}{\lambda^2}
-\frac{(\partial_{\theta}\bar{\lambda})(D_+\bar{\lambda})}{\bar{\lambda}^2}
\nn\\
&\rightarrow&\frac{1}{4}(\partial_{\theta}\sigma)(\partial_t\ln\lambda^+)
+\frac{1}{4}(\partial_{t}\sigma)(\partial_{\theta}\ln\lambda^+)
-\frac{1}{2}(\partial_{\theta}\sigma)(\partial_{\theta}\ln\lambda^+)
\nn\\
&&
-\frac{1}{4}(\partial_{\theta}\sigma)(\partial_t\ln\lambda^-)
-\frac{1}{4}(\partial_{t}\sigma)(\partial_{\theta}\ln\lambda^-)
-\frac{1}{2}(\partial_{\theta}\sigma)(\partial_{\theta}\ln\lambda^-)
\nn\\
&&
-\frac{1}{4}(\partial_{\theta}\sigma)(\partial_{\theta}\sigma);
\eea
\bea
&&
\lambda^2(\partial_{\theta}F)(D_-\Psi)-\bar{\lambda}^2(\partial_{\theta}\bar{F})(D_+\bar{\Psi})
\nn\\
&=&
\frac{1}{4}(\partial_t\sigma)(\partial_{\theta}\ln{\cal F}^+)-\frac{1}{4}(\partial_{\theta}\sigma)(\partial_{\theta}\ln{\cal F}^+)
\nn\\
&&
-\frac{1}{4}(\partial_t\sigma)(\partial_{\theta}\ln{\cal F}^-)-\frac{1}{4}(\partial_{\theta}\sigma)(\partial_{\theta}\ln{\cal F}^-)
\nn\\
&&-\frac{1}{2}(\partial_t\sigma)(\partial_t\sigma)+\frac{1}{2}(\partial_{\theta}\sigma)(\partial_{\theta}\sigma)
+\frac{1}{2}(\partial_t^2\sigma)-\frac{1}{2}(\partial_{\theta}^2\sigma), 
\eea
where 
\bea
{\cal F}^+\equiv\frac{\partial_{\theta}F}{E_{\theta}^+}, \ 
{\cal F}^-\equiv\frac{\partial_{\theta}\bar{F}}{E_{\theta}^-}. 
\eea
\\

\noindent
Because we consider the torus manifold, we can do the integration by part for each direction, and the coupling term between the background and dynamical fields vanishes through the integration by part \cite{Huang:2023aqz}: 
\bea
&&
\frac{k}{\pi}\int dt d\theta\ \bigg(\frac{1}{4}(\partial_{\theta}\sigma)(\partial_t\ln\lambda^+)
+\frac{1}{4}(\partial_{t}\sigma)(\partial_{\theta}\ln\lambda^+)
\nn\\
&&
-\frac{1}{2}(\partial_{\theta}\sigma)(\partial_{\theta}\ln\lambda^+)
\nn\\
&&
+\frac{1}{4}(\partial_t\sigma)(\partial_{\theta}\ln{\cal F}^+)-\frac{1}{4}(\partial_{\theta}\sigma)(\partial_{\theta}\ln{\cal F}^+)
\nn\\
&&
-\frac{1}{4}(\partial_{\theta}\sigma)(\partial_t\ln\lambda^-)
-\frac{1}{4}(\partial_{t}\sigma)(\partial_{\theta}\ln\lambda^-)
\nn\\
&&
-\frac{1}{2}(\partial_{\theta}\sigma)(\partial_{\theta}\ln\lambda^-)
\nn\\
&&
-\frac{1}{4}(\partial_t\sigma)(\partial_{\theta}\ln{\cal F}^-)-\frac{1}{4}(\partial_{\theta}\sigma)(\partial_{\theta}\ln{\cal F}^-)
\bigg)
\nn\\
&=&\frac{k}{\pi}\int dt d\theta\ \bigg(\frac{1}{2}(\partial_{\theta}\sigma)(\partial_t\ln\lambda^+)
-\frac{1}{2}(\partial_{\theta}\sigma)(\partial_{\theta}\ln\lambda^+)
\nn\\
&&
+\frac{1}{4}(\partial_t\sigma)(\partial_{\theta}\ln{\cal F}^+)-\frac{1}{4}(\partial_{\theta}\sigma)(\partial_{\theta}\ln{\cal F}^+)
\nn\\
&&
-\frac{1}{2}(\partial_{\theta}\sigma)(\partial_t\ln\lambda^-)
-\frac{1}{2}(\partial_{\theta}\sigma)(\partial_{\theta}\ln\lambda^-)
\nn\\
&&
-\frac{1}{4}(\partial_t\sigma)(\partial_{\theta}\ln{\cal F}^-)-\frac{1}{4}(\partial_{\theta}\sigma)(\partial_{\theta}\ln{\cal F}^-)
\bigg)
\nn\\
&=&\int dtd\theta\ \bigg\lbrack\frac{1}{4}(\partial_t\sigma)\bigg(\partial_{\theta}\ln\big({\cal F}^+(\lambda^+)^2\big)\bigg)
\nn\\
&&
-\frac{1}{4}(\partial_{\theta}\sigma)\bigg(\partial_{\theta}\ln\big({\cal F}^+(\lambda^+)^2\big)\bigg)
\nn\\
&&
-\frac{1}{4}(\partial_t\sigma)\bigg(\partial_{\theta}\ln\big({\cal F}^-(\lambda^-)^2\big)\bigg)
\nn\\
&&
-\frac{1}{4}(\partial_{\theta}\sigma)\bigg(\partial_{\theta}\ln\big({\cal F}^-(\lambda^-)^2\big)\bigg)
\bigg\rbrack
\nn\\
&=&
0.  
\eea
We use: 
\bea
{\cal F}^+(\lambda^+)^2={\cal F}^-(\lambda^-)^2=r
\eea
in the last equality. 
Hence the Weyl transformation generates the additional term from Eq. \eqref{WZ} as that \cite{Huang:2023aqz}
\bea
\delta S_{\mathrm{B}}=\frac{k}{\pi}\int dtd\theta\ \bigg(
-\frac{1}{2}(\partial_t\sigma)(\partial_t\sigma)
+\frac{1}{4}(\partial_{\theta}\sigma)(\partial_{\theta}\sigma)
\bigg). 
\eea 
The Weyl transformation induces the Liouville theory \cite{Huang:2023aqz}
\bea
\delta S_{\mathrm{B}}+S_{\mathrm{B}1}
=\frac{k}{\pi}\int dtd\theta\ \bigg(-\frac{1}{4}(\partial_t\sigma)(\partial_t\sigma)
+\frac{1}{4}(\partial_{\theta}\sigma)(\partial_{\theta}\sigma)\bigg). 
\eea

\subsection{Partition Function} 
\noindent 
Although the 2D Schwarzian theory is not supersymmetric, we can introduce the non-dynamical fermion field to obtain the supersymmetry \cite{Cotler:2018zff}. 
Because the partition function is one-loop exact, the higher-loop terms do not have the contribution. 
The partition function of 2D Schwarzian theory is 
\bea
Z(\tau)=|q|^{-\frac{C_{\mathrm{cft}_2}}{12}}\frac{1}{\prod^{\infty}_{n=2}|1-q^n|^2}, 
\eea
where 
\bea
q=e^{2\pi i\tau}. 
\eea 
Because the details of the computing partition function overlap with the calculation of EE, we only show the result here. 
With the help of the Dedekind $\eta$ function, 
\bea
\eta(\tau)=q^{\frac{1}{24}}\prod^{\infty}_{n=1}(1-q^n), 
\eea
we simplify the expression of the partition function
\bea
Z(\tau)=\frac{1}{|\eta(\tau)|^2}|q|^{-\frac{1}{12}(C_{\mathrm{cft}_2}-1)}|1-q|^2. 
\eea 
The path integration in the bulk for all asymptotic AdS$_3$ boundary conditions is equivalent to summing over the SL(2, $\mathbb{Z}$) group in the boundary theory \cite{Maloney:2007ud}. 
Due to the summation of all modular transformations, the partition function becomes the manifest modular invariant form \cite{Maloney:2007ud}
\bea
Z_M(\tau)=\sum_{c_1, d_1; (c_1, d_1)=1}Z\bigg(\frac{a_1\tau+b_1}{c_1\tau+d_1}\bigg), 
\eea
where 
\bea
a_1d_1-b_1c_1=1, \ a_1, b_1, c_1, d_1\in \mathbb{Z}.
\eea
The relative prime $c_1$ and $d_1$ implies that
\bea
(c_1, d_1)=1. 
\eea
If we have two solutions for $a_1, b_1$ and $a_2, b_2$, we then obtain that
\bea
(a_1-a_2)d_1=(b_1-b_2)c_1. 
\eea
We can further solve the above equation:
\bea
a_1-a_2=mc_1, \ b_1-b_2=md_1, 
\eea
where $m$ is an arbitrary integer. 
The transformation:
\bea
a_1\rightarrow a_1+mc_1; \ b_1\rightarrow b_1+md_1, 
\eea
implies
\bea
\frac{a_1\tau+b_1}{c_1\tau+d_1}\rightarrow\frac{a_1\tau+b_1}{c_1\tau+d_1}+m.
\eea
and
\bea
q\rightarrow qe^{2\pi im}=q.
\eea
Because the partition function only depends on $q$, we have the redundancy from the different choices of $m$. 
Hence we only sum over $m=0$ in the partition function, which is equivalent to not including $a$ and $b$. 
We will further simplify the expression of $Z_M$ by showing that $\sqrt{\mathrm{Im}(\tau)}|\eta(\tau)|^2$ is the modular invariant \cite{Maloney:2007ud}. 

\subsubsection{Modular Invariant Variable} 
\noindent 
There are two generators that produce the modular transformation: the T-transformation and the S-transformation. 
The T-transformation is:
\bea
\tau\rightarrow \tau+1; \qquad a_1=b_1=d_1=1, \ c_1=0.  
\eea
The S-transformation is: 
\bea
\tau\rightarrow-\frac{1}{\tau}; \qquad 
a_1=d_1=0, \ b_1=-c_1=1. 
\eea
$q$ is invariant under the T-transformation. 
Therefore, $\sqrt{\mathrm{Im}(\tau)}|\eta(\tau)|^2$ is a modular invariant variable. 
\\

\noindent 
For discussing the S-transformation, we first define that 
\bea
r_1\equiv\exp\bigg(-\frac{2\pi i}{\tau}\bigg). 
\eea
We then obtain 
\bea
\eta\bigg(-\frac{1}{\tau}\bigg)=r_1^{\frac{1}{24}}\prod_{n=1}^{\infty}(1-r_1^n). 
\eea
By using the pentagonal number theorem (only when $|r|_1<1$, $\mathrm{Im}(r_1)>0$, or $\mathrm{Im}(\tau)>0$)
\bea
\prod_{n=1}^{\infty}(1-x^n)=\sum_{k=-\infty}^{\infty}(-1)^kx^{\frac{k(3k-1)}{2}}, 
\eea
we can rewrite the infinite product as an infinite summation in the Dedekind $\eta$ function:
\bea
\eta\bigg(-\frac{1}{\tau}\bigg)=r_1^{\frac{1}{24}}\sum_{n=-\infty}^{\infty}(-1)^nr_1^{\frac{(3n^2-n)}{2}}
=\sum_{n=-\infty}^{\infty}f(n), 
\eea
where 
\bea
f(n)\equiv\exp\bigg\lbrack\pi i\bigg(-\frac{1}{12\tau}+n-\frac{3n^2-n}{\tau}\bigg)\bigg\rbrack.
\eea
We then do an analytical continuation from $n$ to the real number $x$ and find the Fourier transform as that: 
\bea
\tilde{f}(k)&=&\int_{-\infty}^{\infty}dx\ \exp\bigg\lbrack\pi i\bigg(-\frac{1}{12\tau}+x-2kx-\frac{3x^2-x}{\tau}\bigg)\bigg\rbrack
\nn\\
&=&\sqrt{\frac{-i\tau}{3}}\exp\bigg\lbrack\pi i\bigg(\frac{\tau(2k-1)^2}{12}-\frac{2k-1}{6}\bigg)\bigg\rbrack. 
\label{AC}
\eea
We use the result of Gaussian integration
\bea
\int_{-\infty}^{\infty}dx\ \exp(-a_2x^2+b_2x+c_2)=\sqrt{\frac{\pi}{a_2}}\exp\bigg(\frac{b_2^2}{4a_2}+c_2\bigg)
\eea
with the choices of the parameters: 
\bea
a_2=\frac{3\pi i}{\tau}, \ b_2=\bigg(1-2k+\frac{1}{\tau}\bigg)\pi i, \ c_2=-\frac{\pi i}{12\tau}.
\eea 
\\

\noindent
The process of analytically continuing from an integer to a real number is not rigorous. 
This can be proven more easily and understandably by following a simpler and more intuitive method. 
Nonetheless, it is possible to obtain rigorous proof without using analytical continuation. 
We then apply Eq. \eqref{AC} to the integer $k$ case to obtain that
\bea
\eta\bigg(-\frac{1}{\tau}\bigg)=\sqrt{\frac{-i\tau}{3}}\sum_{k=-\infty}^{\infty}
\exp\bigg\lbrack\pi i\bigg(\frac{\tau(2k-1)^2}{12}-\frac{2k-1}{6}\bigg)\bigg\rbrack.
\eea
Using the pentagonal number theorem also provides the following similar formula: 
\bea
\eta(\tau)=q^{\frac{1}{24}}\sum_{n=-\infty}^{\infty}(-1)^nq^{\frac{3n^2-n}{2}}
=\sum_{n=-\infty}^{\infty}\bigg\lbrack\pi i\bigg(\frac{\tau(6n-1)^2}{12}+n\bigg)\bigg\rbrack.
\label{DEF1}
\eea
The integer $k$ can be decomposed as $3l, 3l+1, 3l+2$, where $l\in\mathbb{Z}$. 
Therefore, we obtain the following results: 
\bea
&&
\eta\bigg(-\frac{1}{\tau}\bigg)
\nn\\
&=&\sqrt{\frac{-i\tau}{3}}
\nn\\
&&\times
\sum_{l=-\infty}^{\infty}\Bigg\lbrack
\exp\bigg\lbrack\pi i\bigg(\frac{\tau(6l-1)^2}{12}-\frac{6l-1}{6}\bigg)\bigg\rbrack
\nn\\
&&+\exp\bigg\lbrack\pi i\bigg(\frac{\tau(6l+1)^2}{12}-\frac{6l+1}{6}\bigg)\bigg\rbrack
\nn\\
&&+\exp\bigg\lbrack\pi i\bigg(\frac{\tau(6l+3)^2}{12}-\frac{6l+3}{6}\bigg)\bigg\rbrack
\Bigg\rbrack
\nn\\
&=&\sqrt{\frac{-i\tau}{3}}
\nn\\
&&\times
\sum_{l=-\infty}^{\infty}\Bigg\lbrack
\exp\bigg\lbrack\pi i\bigg(\frac{\tau(6l-1)^2}{12}-l\bigg)\bigg\rbrack\exp\bigg(\frac{\pi i}{6}\bigg)
\nn\\
&&+\exp\bigg\lbrack\pi i\bigg(\frac{\tau(6l-1)^2}{12}+l\bigg)\bigg\rbrack\exp\bigg(-\frac{\pi i}{6}\bigg)
\nn\\
&&+\exp\bigg\lbrack\pi i\bigg(\frac{\tau(6l+3)^2}{12}-\frac{6l+3}{6}\bigg)\bigg\rbrack
\Bigg\rbrack
\nn\\
&=&\sqrt{-i\tau}
\nn\\
&&\times
\sum_{l=-\infty}^{\infty}\Bigg\lbrack
\exp\bigg\lbrack\pi i\bigg(\frac{\tau(6l-1)^2}{12}-l\bigg)\bigg\rbrack
\nn\\
&&+\exp\bigg\lbrack\pi i\bigg(\frac{\tau(6l+3)^2}{12}-\frac{6l+3}{6}\bigg)\bigg\rbrack
\Bigg\rbrack. 
\label{DEF}
\eea
In the last equality of Eq. \eqref{DEF}, we use the further simplification:
\bea
&&
\sum_{l=-\infty}^{\infty}\exp\bigg\lbrack\pi i\bigg(\frac{\tau(6l+3)^2}{12}-\frac{6l+3}{6}\bigg)\bigg\rbrack
\nn\\
&=&
\sum_{m\in 2\mathbb{Z}+1}\exp\bigg\lbrack\pi i\bigg(\frac{3\tau m^2}{4}-\frac{m}{2}\bigg)\bigg\rbrack
\nn\\
&=&\sum_{m\in2\mathbb{Z}+1, m>0}\exp\bigg(\pi i\frac{3\tau m^2}{4}\bigg)\bigg(e^{\frac{\pi im}{2}}+e^{\frac{-\pi im}{2}}\bigg)
\nn\\
&=&0,  
\eea
in which we use: 
\bea
e^{\pi im}=-1, \ m\in2\mathbb{Z}+1 
\eea
in the last equality. 
Combining Eqs. \eqref{DEF} and \eqref{DEF1} shows that:
\bea
\eta\bigg(-\frac{1}{\tau}\bigg)=\sqrt{-i\tau}
\sum_{l=-\infty}^{\infty}\Bigg\lbrack
\exp\bigg\lbrack\pi i\bigg(\frac{\tau(6l-1)^2}{12}-l\bigg)\bigg\rbrack=\sqrt{-i\tau}\eta(\tau).
\eea
The modular invariance property can be demonstrated in the following manner:  
\bea
\sqrt{\mathrm{Im}\bigg(-\frac{1}{\tau}\bigg)}\bigg|\eta\bigg(-\frac{1}{\tau}\bigg)\bigg|^2
=\sqrt{\mathrm{Im}\bigg(-\frac{1}{\tau}\bigg)|\tau|^2}|\eta(\tau)|^2
=\sqrt{\mathrm{Im}(\tau)}|\eta(\tau)|^2.
\nn\\
\eea
Therefore, we can simplify the partition function even further 
\bea
Z_M(\tau)=\frac{1}{\sqrt{\mathrm{Im}(\tau)}|\eta(\tau)|^2}
\sum_{c, d}(\sqrt{\mathrm{Im}(\tau)}|q|^{-\frac{1}{12}(C_{\mathrm{cft}_2}-1)}|1-q|^2)|_{M}, 
\eea
where $(\cdots)|_M$ means that 
\bea
\tau\rightarrow\frac{a_1\tau+b_1}{c_1\tau+d_1}, 
\eea 
when $\mathrm{Im}(\tau)>0$.

\subsubsection{Analytical Expression} 
\noindent 
We will expand $|1-q|^2$ and express the $Z(\tau)$ as a sum of the Poincaré series \cite{Maloney:2007ud}
\bea
P(\tau; n, m)\equiv\sum_{c_1, d_1; (c_1, d_1)=1}(\sqrt{\mathrm{Im}\tau}q^{-n}\bar{q}^{-m})|_M.
\eea
In general, this series is divergent. 
We need to regularize the series through the analytical continuation \cite{Maloney:2007ud}
\bea
P_1(\tau; s, n, m)\equiv\sum_{c_1, d_1; (c_1, d_1)=1}\big((\mathrm{Im}\tau)^sq^{-n}\bar{q}^{-m}\big)\big|_M.
\eea
In the domain $\mathrm{Re}(s)>1$, we can get the following result 
\bea
&&
P_1(\tau; s, n, m)
\nn\\
&=&y_1^se^{2\pi(\kappa y_1+i\mu x_1)}
\nn\\
&&
+\sum_{c_1>0, d_1}\frac{y_1^s}{|c_1\tau+d_1|^2}\exp\bigg\lbrack\frac{2\pi\kappa y_1}{|c_1\tau+d_1|^2}
+2\pi i\mu\bigg(\frac{a_1}{c_1}-\frac{c_1x_1+d_1}{c_1|c_1\tau+d_1|^2}\bigg)\bigg\rbrack, 
\nn\\
\label{P1}
\eea
where 
\bea
\kappa\equiv m+n, \ \mu\equiv m-n, \ \tau\equiv x_1+iy_1,  
\eea
and then do the analytical continuation to $s=1/2$. 
The first term of Eq. \eqref{P1} is given by $c_1=0$. 
Because $(0, d_1)=d_1$, we choose $d_1=1$ in the first term. 
The following fact is helpful for the calculation: 
\bea
\mathrm{Re}\bigg(\frac{a_1\tau+b_1}{c_1\tau+d_1}\bigg)&=&\frac{a_1c_1|\tau|^2+b_1d_1+(a_1d_1+b_1c_1)x_1}{|c_1\tau+d_1|^2}, 
\nn\\
\mathrm{Im}\bigg(\frac{a_1\tau+b_1}{c_1\tau+d_1}\bigg)&=&\frac{y_1}{|c_1\tau+d_1|^2}. 
\nn\\
\eea 
In the end, we obtain the analytical expression of the partition function after the analytical continuation: 
\bea
&&
Z_M(\tau)
\nn\\
&=&\frac{1}{\sqrt{\mathrm{Im}(\tau)}|\eta(\tau)|^2}
\nn\\
&&\times\bigg\lbrack
P\bigg(\tau; \frac{C_{\mathrm{cft}_2}-1}{24}, \frac{C_{\mathrm{cft}_2}-1}{24}\bigg)
\nn\\
&&
-P\bigg(\tau; \frac{C_{\mathrm{cft}_2}-1}{24}-1, \frac{C_{\mathrm{cft}_2}-1}{24}\bigg)
\nn\\
&&
-P\bigg(\tau; \frac{C_{\mathrm{cft}_2}-1}{24}, \frac{C_{\mathrm{cft}_2}-1}{24}-1\bigg)
\nn\\
&&
+P\bigg(\tau; \frac{C_{\mathrm{cft}_2}-1}{24}-1, \frac{C_{\mathrm{cft}_2}-1}{24}-1\bigg)\bigg\rbrack
\nn\\
&\sim&\frac{1}{\sqrt{\mathrm{Im}(\tau)}|\eta(\tau)|^2}
\nn\\
&&\times\bigg\lbrack
P_1\bigg(\tau; \frac{C_{\mathrm{cft}_2}-1}{24}, \frac{C_{\mathrm{cft}_2}-1}{24}\bigg)
\nn\\
&&
-P_1\bigg(\tau; \frac{C_{\mathrm{cft}_2}-1}{24}-1, \frac{C_{\mathrm{cft}_2}-1}{24}\bigg)
\nn\\
&&
-P_1\bigg(\tau; \frac{C_{\mathrm{cft}_2}-1}{24}, \frac{C_{\mathrm{cft}_2}-1}{24}-1\bigg)
\nn\\
&&
+P_1\bigg(\tau; \frac{C_{\mathrm{cft}_2}-1}{24}-1, \frac{C_{\mathrm{cft}_2}-1}{24}-1\bigg)\bigg\rbrack.
\nn\\
\eea

\section{EE and AdS$_3$/CFT$_2$ Correspondence}
\label{sec:4}
\noindent
To introduce EE, we begin with a discussion of classical Shannon entropy. 
We then introduce a reduced density matrix to define EE. 
Since the loss of a reduced density matrix in QFT, it is necessary to use the replica trick \cite{Holzhey:1994we} to compute. 
We review the procedure of the replica trick. 
In the end, we review the holographic EE \cite{Ryu:2006bv, Ryu:2006ef} and the dual of EE from bulk Wilson line \cite{Ammon:2013hba,deBoer:2013vca,Huang:2019nfm}. 

\subsection{EE}
\noindent
We discuss the quantification of received information in the context of information theory, introducing a function $S(p)$ dependent on probability $p$. 
When one knows that the probability is one, it implies no surprise. 
Therefore, the $S$ is 
\bea
S(1)=0.
\eea
The surprising level decreases as the probability increases: 
\bea
S(p)>S(q), \ p<q.
\eea
When two events are independent, the $S$ has the additive property
\bea
S(pq)=S(p)+S(q). 
\eea
Finally, we assume that the $S$ is a continuous function of the probability.
\\ 

\noindent
Using the additive property and the continuity of the $S$ shows that 
\bea
S(p^x)=xS(p)
\eea
with $x$ as a positive rational number. 
Choosing 
\bea
x=-\ln p,
\eea
where 
\bea
0<p\le 1,
\eea
leads to
\bea
S(p)=S(e^{-x})=x\cdot S\bigg(\frac{1}{e}\bigg)=-C\cdot\ln p,
\eea
where $C$ is a constant, 
\bea
C\equiv S\bigg(\frac{1}{e}\bigg)>S(1)=0.
\eea 
Therefore, we can derive a unique function that accurately defines the reception of information. 
The classical Shannon entropy is the expectation value of the $S$
\bea
S_{c}\equiv-\sum_j p_j\ln p_j.
\eea
\\

\noindent
When defining the von Neumann entropy, we replace the probability with a density matrix. 
For a pure quantum state in regions $A$ and $B$, the density matrix is 
\bea
\rho_{AB}\equiv|\psi\rangle\langle\psi|,
\eea
where $|\psi\rangle$ is a quantum state. 
The density matrix has already been normalized as
\bea
\mathrm{Tr}(\rho_{AB})=1, 
\eea
which means that the trace of the density matrix is equal to one. 
The von Neumann entropy for the entire system is
\bea
S_{vn}\equiv-\mathrm{Tr}(\rho_{AB}\ln\rho_{AB}).
\eea
When one partial traces over a region $B$, one obtains a reduced density matrix for the region $A$
\bea
\rho_A\equiv\mathrm{Tr}_B(\rho_{AB}).
\eea 
The EE for region $A$ is
\bea
S_{EE, A}\equiv-\mathrm{Tr}(\rho_A\ln\rho_A).
\eea

\subsection{Replica Trick} 
\begin{figure}[!htb]
\begin{center}
\includegraphics[width=1.\textwidth]{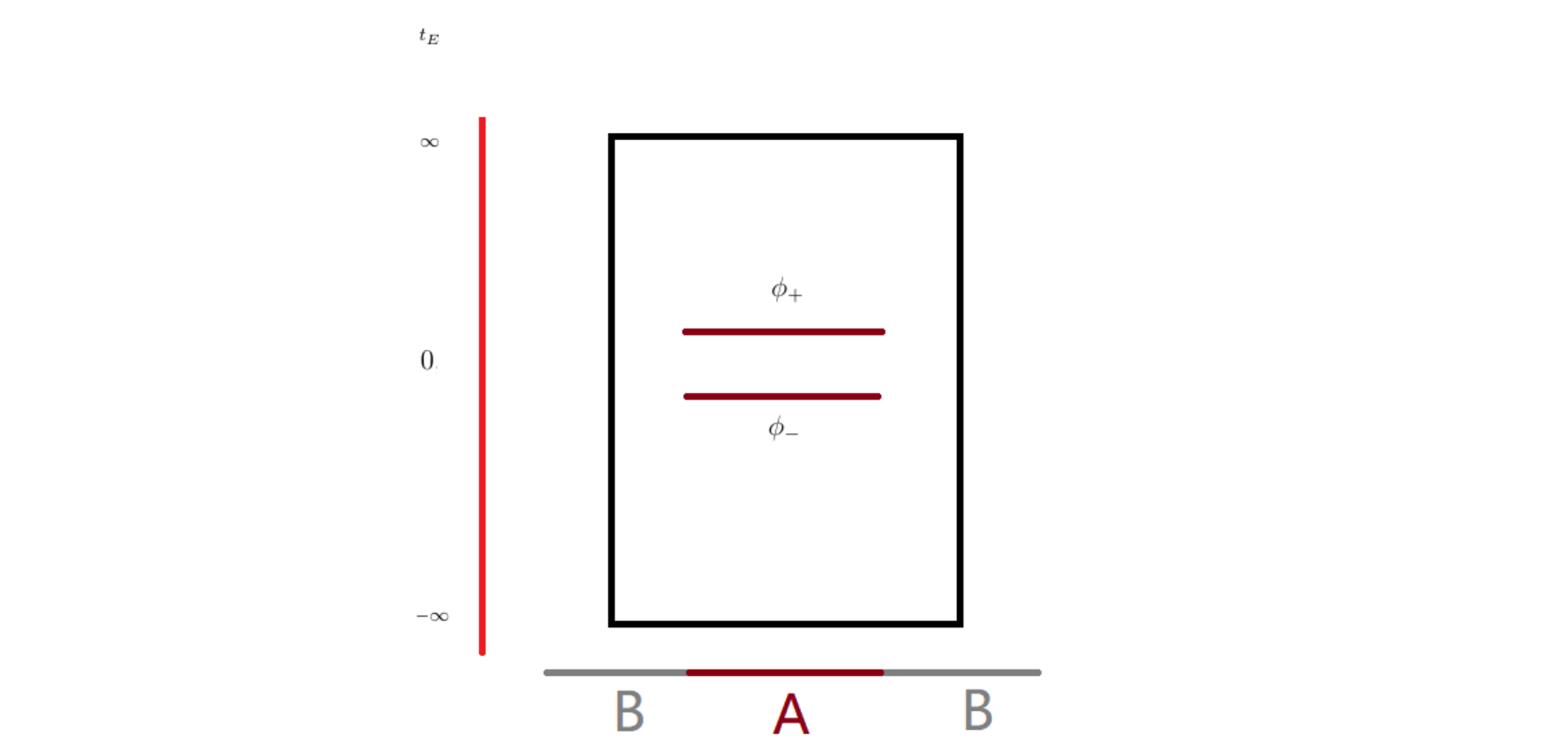}
\caption{The path integration of a reduced density matrix $\rho_{AB}$.}
\label{region.pdf}
\end{center}
\end{figure}

\noindent 
We first introduce the field theory technique, replica trick \cite{Holzhey:1994we}, by the single interval case as in Fig. \ref{region.pdf}. 
The ground state wavefunctional is given by 
\bea
\psi\big(\phi_0(\vec{x})\big)=\int_{t_E=-\infty}^{\phi(t_E=0, \vec{x})=\phi_0(\vec{x})}{\cal D}\phi\ e^{-S(\phi)}, 
\eea 
where the flat Euclidean coordinates are $(t_E, \vec{x})$, and $S(\phi)$ is an action of a physical system. 
\\

\noindent
A density matrix $\rho_{AB}$ is given by 
\bea
(\rho)_{\phi_0\tilde{\phi}_0}
=
\psi(\phi_0)\phi^{\dagger}(\tilde{\phi}_0). 
\eea
The complex conjugate one $\psi^{\dagger}$ is given by a path integration from $t_E=\infty$ to $t_E=0$ 
\bea
\psi^{\dagger}\big(\tilde{\phi}_0\big)=\int_{t_E=\infty}^{\tilde{\phi}_0}{\cal D}\phi\ e^{-S(\phi)}. 
\eea 
A qubit state is a linear superposition of the different orthogonal states. 
Here $\phi_0$ plays a similar role to the orthogonal state. 
\\ 

\noindent 
We integrate our $\phi_0$ on the region $B$ with that
\bea
\phi_0=\tilde{\phi}_0 
\label{ptc}
\eea
for obtaining a reduced density matrix of the region $A$ 
\bea
&&
(\rho_A)_{\phi_+\phi_-}
\nn\\
&=&
\frac{1}{Z_1}\int_{t_E=-\infty}^{t_E=\infty}{\cal D}\phi\ e^{-S(\phi)}\prod_{\vec{x}\in A} 
\delta\big(\phi(0^+, \vec{x})-\phi_+(0, \vec{x})\big) 
\delta\big(\phi(0^-, \vec{x})-\phi_-(0, \vec{x})\big), 
\nn\\
\eea
where $Z_1$ is a partition function for a normalization 
\bea
\mathrm{Tr}_A\rho_A=1. 
\eea 
Performing a partial trace operation on one qubit results in the same level number for the orthogonal states of a qubit state. 
The condition \eqref{ptc} is similar to imposing the same level number. 
\\

\begin{figure}[!htb]
\begin{center}
\includegraphics[width=1.\textwidth]{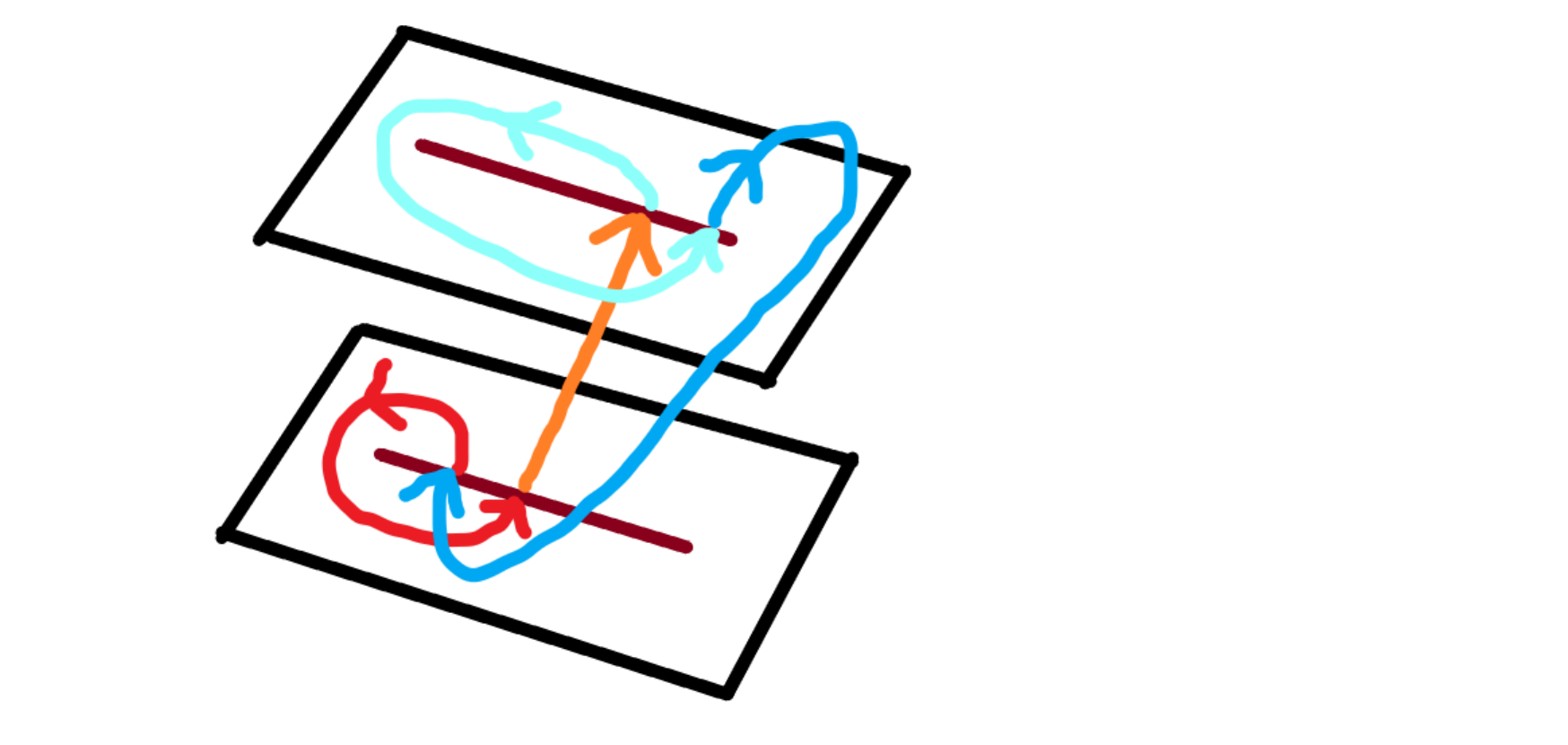}
\caption{The demonstration of $n$-sheet manifold ${\cal R}_n$ for $n=2$.}
\label{rt.pdf}
\end{center}
\end{figure}

\noindent
Because it is hard to compute with the singularity at $t_E=0$, we prepare $n$ copies of the reduced density matrix of the region $A$ as that 
\bea
(\rho_A)_{\phi_{1+}\phi_{1-}}(\rho_A)_{\phi_{2+}\phi_{2-}}\cdots (\rho_A)_{\phi_{n+}\phi_{n-}} 
\eea 
with a boundary condition 
\bea
\phi_{j-}(\vec{x})=\phi_{(j+1)+}(\vec{x}), \ j=1, 2, \cdots, n, 
\eea
where 
\bea
\phi_{(n+1)+}(\vec{x})\equiv \phi_{1+}(\vec{x}). 
\eea 
We then integrate each $\phi_{j+}$ to realize a partial trace operation. 
Therefore, a path-integral representation of $\mathrm{Tr}_A\rho_A^n$ can be provided: 
\bea
\mathrm{Tr}_A\rho_A^n
= 
\frac{1}{Z_1^n}\int _{(t_E, \vec{x})\in {\cal R}_n} {\cal D}\phi\ e^{-S(\phi)}\equiv \frac{Z_n}{Z_1^n}, 
\eea
where ${\cal R}_n$ is an $n$-sheet manifold as in Fig. \ref{rt.pdf}, and $Z_n$ is an $n$-sheet partition function (defined on an ${\cal R}_n$). 
In the end, we can take the one-sheet limit $n\rightarrow 1$
on Rényi entropy 
\bea
S_n\equiv \frac{\ln\mathrm{Tr}_A\rho_A^n}{1-n}
\eea
to obtain EE as that
\bea
S_1=\lim_{n\rightarrow1}S_n. 
\eea

\subsection{RT Conjecture}
\noindent
We introduce the RT conjecture \cite{Ryu:2006bv, Ryu:2006ef} in this section. 
The RT conjecture states that an AdS$_{d+1}$ bulk minimum surface is dual to EE of CFT$_d$. 
We demonstrate the AdS$_3$ case. 

\noindent 
The AdS$_3$ spacetime in the Poincaré coordinate is given by
\bea
ds_{3P}^2=-\frac{1}{\Lambda}\frac{dt^2+dx^2+dz^2}{z^2}.
\eea 
The AdS$_3$ induced metric is given by:
\bea
ds_{3b}^2=h_{\mu\nu}dx^{\mu}dx^{\nu}=-\frac{1}{\Lambda}\frac{1}{z^2}\bigg\lbrack 1+\bigg(\frac{dz}{dx}\bigg)^2\bigg\rbrack dx^2
\eea
by choosing a time slice ($t$ as a constant). 
Hence the area of this surface is given by
\bea
A_{AdS_3}=\sqrt{-\frac{1}{\Lambda}}\int dx\ \frac{1}{z}\sqrt{ 1+\bigg(\frac{dz}{dx}\bigg)^2}.
\eea
\\

\noindent
The minimum area satisfies the relation:
\bea
\frac{d}{dx}\frac{\delta A_{\mathrm{AdS}_3}}{\delta z^{\prime}}=\frac{\delta A_{\mathrm{AdS}_3}}{\delta z}, \ 
\frac{d}{dx}\bigg\lbrack\frac{\frac{dz}{dx}}{z}\frac{1}{\sqrt{1+\big(\frac{dz}{dx}\big)^2}}\bigg\rbrack=-\frac{1}{z^2}\sqrt{1+\bigg(\frac{dz}{dx}\bigg)^2},
\eea
where 
\bea
z^{\prime}\equiv \frac{dz}{dx}.
\eea 
One solution is:
\bea
z(x)=\sqrt{L^2-x^2}, \ \frac{dz}{dx}=-\frac{x}{z}.
\eea
We check the solution as in the following:
\bea
-\frac{1}{z^2}\sqrt{1+\bigg(\frac{dz}{dx}\bigg)^2}&=&-\frac{1}{z^2}\sqrt{1+\frac{x^2}{z^2}}=-\frac{L}{(L^2-x^2)^{\frac{3}{2}}},
\nn\\
\frac{d}{dx}\bigg\lbrack\frac{\frac{dz}{dx}}{z}\frac{1}{\sqrt{1+\bigg(\frac{dz}{dx}\bigg)^2}}\bigg\rbrack&=&\frac{d}{dx}\bigg\lbrack-\frac{x}{z^2}\frac{1}{\sqrt{1+\frac{x^2}{z^2}}}\bigg\rbrack
\nn\\
&=&-\frac{d}{dx}\bigg(\frac{x}{z\sqrt{z^2+x^2}}\bigg)=-\frac{d}{dx}\bigg(\frac{x}{Lz}\bigg)
\nn\\
&=&-\frac{1}{Lz}-\frac{x^2}{Lz^3}=-\frac{1}{L}\frac{z^2+x^2}{z^3}=-\frac{L}{z^3}
\nn\\
&=&-\frac{L}{(L^2-x^2)^{\frac{3}{2}}}=-\frac{1}{z^2}\sqrt{1+\bigg(\frac{dz}{dx}\bigg)^2}.
\eea
\\

\noindent
The minimum area is given by:
\bea
A_{\mathrm{AdS}_3}&=&\sqrt{-\frac{1}{\Lambda}}\int_{-L+\delta}^{L-\delta} dx\ \frac{1}{z}\sqrt{ 1+\bigg(\frac{dz}{dx}\bigg)^2}=\sqrt{-\frac{1}{\Lambda}}\int_{-L+\delta}^{L-\delta} dx\ \frac{1}{z}\sqrt{ 1+\bigg(\frac{x}{z}\bigg)^2}
\nn\\
&=&\sqrt{-\frac{1}{\Lambda}}\int_{-L+\delta}^{L-\delta} dx\ \frac{L}{z^2}=\sqrt{-\frac{1}{\Lambda}}\int_{-L+\delta}^{L-\delta} dx\ \frac{L}{L^2-x^2}
\nn\\
&=&\frac{1}{2}\sqrt{-\frac{1}{\Lambda}}\int_{-L+\delta}^{L-\delta} dx\ \bigg(\frac{1}{L-x}+\frac{1}{L+x}\bigg)
\nn\\
&=&\sqrt{-\frac{1}{\Lambda}}\int_{0}^{L-\delta} dx\ \bigg(\frac{1}{L-x}+\frac{1}{L+x}\bigg)
\nn\\
&=&\sqrt{-\frac{1}{\Lambda}}\ln\bigg|\frac{L+x}{x-L}\bigg|\Bigg|^{L-\delta}_{0}=\sqrt{-\frac{1}{\Lambda}}\ln\frac{2L-\delta}{\delta}.
\eea
The integration range excludes the boundary sites $L$ and $-L$ corresponding to CFT$_2$'s entangling surface. 
We then define that:
\bea
\epsilon\equiv\sqrt{L^2-(L-\delta)^2}=\sqrt{2\delta L-\delta^2}.
\eea
Hence we can obtain two solutions
\bea
\delta=L\pm\sqrt{L^2-\epsilon^2}.
\eea
If we assume $L\gg\epsilon>0$, we should choose the solution
\bea
\delta=L-\sqrt{L^2-\epsilon^2}.
\eea
Now minimum area becomes:
\bea
A_{AdS_3}=\sqrt{-\frac{1}{\Lambda}}\ln\frac{2L-\delta}{\delta}=\sqrt{-\frac{1}{\Lambda}}\ln\frac{L+\sqrt{L^2-\epsilon^2}}{L-\sqrt{L^2-\epsilon^2}}.
\eea
The holographic entanglement entropy for the AdS$_3$ metric is shown as that \cite{Ryu:2006bv, Ryu:2006ef}:
\bea
\frac{A_{\mathrm{AdS}_3}}{4G_3}&=&\frac{1}{4\sqrt{-\Lambda}G_3}\ln\frac{L+\sqrt{L^2-\epsilon^2}}{L-\sqrt{L^2-\epsilon^2}}
\nn\\
&=&\frac{c_{\mathrm{cft}_2}}{6}\ln\frac{L+\sqrt{L^2-\epsilon^2}}{L-\sqrt{L^2-\epsilon^2}}=\frac{c_{\mathrm{cft}_2}}{6}\ln\frac{4L^2}{\epsilon^2}+\cdots
\nn\\
&=&\frac{c_{\mathrm{cft}_2}}{3}\ln\frac{2L}{\epsilon}+\cdots
=\frac{c_{\mathrm{cft}_2}}{3}\ln\frac{L}{\epsilon}+\cdots,
\label{heeads3}
\eea
the center charge of CFT$_2$ is given by \cite{Brown:1986nw}
\bea
c_{\mathrm{cft}_2}\equiv\frac{3}{2\sqrt{-\Lambda}G_3}
\eea
We use the following expansion:
\bea
&&
L+\sqrt{L^2-\epsilon^2}=2L+\cdots, 
\nn\\
&& 
L-\sqrt{L^2-\epsilon^2}=L\bigg(1-\frac{\epsilon^2}{2L^2}\bigg)+\cdots=L-\frac{\epsilon^2}{2L}
\eea
in the third equality of Eq. \eqref{heeads3}. 
Therefore, we have demonstrated that the RT conjecture is capable of reproducing the entanglement entropy of CFT$_2$.

\subsection{EE in 2D Schwarzian Theory}
\noindent 
We apply the conformal mapping to transform a planar to a two-sphere ($S^2$) \cite{Casini:2011kv}. 
The removal of the top and bottom points from $S^2$ results in a cylinder manifold. 
This manifold serves as the asymptotic boundary for the Lorentzian AdS$_3$ in global coordinates. 
Once we identify the boundary and apply the Wick rotation to the AdS$_3$ spacetime, the cylinder transforms into a torus. 
This allows us to calculate the EE using the partition function of the $n$-sheet torus \cite{Huang:2019nfm,Huang:2023aqz}. 
However, this method may not account for the global effect resulting from the boundary identification. 

\subsubsection{Conformal Mapping} 
\noindent
The planar manifold is
\bea
ds_P^2=dt^2+dx^2,
\eea 
where 
\bea
-\infty< t, x<\infty. 
\eea
The entangling surface, a boundary of two subregions, is at $(t, |x|)=(0, L)$. 
The length of the interval is $2L$ at $t=0$. 
We define new coordinate variables $\tau$ and $u$:
\bea
t&=&L\frac{\sin\big(\frac{\tau}{L}\big)}{\cosh\big(u\big)+\cos\big(\frac{\tau}{L}\big)}; \qquad x=L\frac{\sinh\big(u\big)}{\cosh\big( u\big)+\cos\big(\frac{\tau}{L}\big)}, 
\eea
where
\bea
-\infty< u <\infty; \ 0\le\frac{\tau}{L}< 2\pi n.
\eea
on the $n$-sheet manifold, which gives the following terms:
\bea
dt
&=&\frac{\bigg(1+\cos\big(\frac{\tau}{L}\big)\cosh\big(u\big)\bigg)d\tau}{\bigg(\cosh\big(u)+\cos\big(\frac{\tau}{L}\big)\bigg)^2}
-\frac{L\sin\big(\frac{\tau}{L}\big)\sinh\big(u\big)du}{\bigg(\cosh\big( u\big)+\cos\big(\frac{\tau}{L}\big)\bigg)^2},
\nn\\
dx
&=&\frac{L\bigg(1+\cos\big(\frac{\tau}{L}\big)\cosh\big(u\big)\bigg)du}{\bigg(\cosh\big(u)+\cos\big(\frac{\tau}{L}\big)\bigg)^2}
+\frac{\sin\big(\frac{\tau}{L}\big)\sinh\big(u\big)d\tau}{\bigg(\cosh\big( u\big)+\cos\big(\frac{\tau}{L}\big)\bigg)^2},
\nn\\
dt^2+dx^2
&=&\frac{d\tau^2}{\bigg(\cosh^2\big(u\big)+\cos^2\big(\frac{\tau}{L}\big)\bigg)^2}
+L^2\frac{du^2}{\bigg(\cosh^2\big(u\big)+\cos^2\big(\frac{\tau}{L}\big)\bigg)^2},
\nn\\
\eea
and then we obtain the metric in the new coordinate as that
\bea
ds_P^2
=\frac{d\tau^2}{\bigg(\cosh^2\big(u\big)+\cos^2\big(\frac{\tau}{L}\big)\bigg)^2}
+L^2\frac{du^2}{\bigg(\cosh^2\big(u\big)+\cos^2\big(\frac{\tau}{L}\big)\bigg)^2}.
\eea
In CFT, we can omit the common pre-factor, which generates the Liouville theory, and the new metric becomes
\bea
ds_{P1}^2 = \frac{d\tau^2}{L^2}+du^2.
\eea 
The CFT partition function is invariant under a local rescaling of the metric (or the Weyl transformation). 
We then redefine 
\bea
\sinh\big(u\big)=\cot\big(\psi\big),
\eea 
where 
\bea
0\le\psi <\pi,
\eea
 and get
\bea
ds_{P1}^2 = \frac{d\tau^2}{L^2}+\frac{d\psi^2}{\sin^2\big(\psi\big)},
\eea
in which we use
\bea
du^2=\frac{d\psi^2}{\sin^2\big(\psi\big)}.
\eea
We omit the pre-factor to obtain the sphere manifold \cite{Casini:2011kv}
\bea
ds_{P2}^2 = d\psi^2+\sin^2\big(\psi\big)\frac{d\tau^2}{L^2}.
\eea 
\\

\noindent
The sphere manifold is isomorphic to a cylinder manifold by removing the top and bottom points. 
We can use: 
\bea
\sech(y)\equiv\sin(\psi); \ d\theta\equiv\frac{d\tau}{L}
\eea
to obtain 
\bea
ds_{P2}^2 = \sech^2(y)(dy^2+d\theta^2). 
\eea 
After removing the pre-factor $\sech(y)$, we obtain the cylinder manifold 
\bea
ds_{C}^2=dy^2+d\theta^2,   
\eea
where 
\bea
0\le \theta<2\pi n. 
\eea 
To compute the $n$-sheet partition function, we regularize the range of the $y$-direction 
\bea
-\ln\bigg(\frac{L}{\epsilon}\bigg)<y<\ln\bigg(\frac{L}{\epsilon}\bigg), 
\eea 
where $\epsilon$ represents a number infinitely close to zero, and also identifies the boundary:
\bea
z_n\equiv\frac{\theta+iy}{n}\sim z_n+2\pi\tau_n, 
\eea 
where $\tau_n$ is the complex structure of the $n$-sheet torus 
\bea
\tau_n=\frac{i}{n\pi}\ln\bigg(\frac{L}{\epsilon}\bigg). 
\eea 
 to obtain the torus manifold. 
 The boundary fields in Eq. \eqref{OE} satisfy the boundary conditions: 
 \bea
 &&
 \phi\bigg(\frac{y}{n}, \frac{\theta}{n}+2\pi\bigg)=\phi\bigg(\frac{y}{n}, \frac{\theta}{n}\bigg)+2\pi, 
 \nn\\
 && 
 \phi\bigg(\frac{y}{n}+2\pi\cdot\mathrm{Im}(\tau_n), \frac{\theta}{n}+2\pi\cdot\mathrm{Re}(\tau_n)\bigg)=\phi\bigg(\frac{y}{n}, \frac{\theta}{n}\bigg); 
 \nn\\
  &&
 \bar{\phi}\bigg(\frac{y}{n}, \frac{\theta}{n}+2\pi\bigg)=\bar{\phi}\bigg(\frac{y}{n}, \frac{\theta}{n}\bigg)+2\pi, 
 \nn\\
 && 
 \bar{\phi}\bigg(\frac{y}{n}+2\pi\cdot\mathrm{Im}(\tau_n), \frac{\theta}{n}+2\pi\cdot\mathrm{Re}(\tau_n)\bigg)=\bar{\phi}\bigg(\frac{y}{n}, \frac{\theta}{n}\bigg).
 \eea

\subsubsection{EE for a Single Interval}
\noindent
We calculate the R\'enyi entropy 
\begin{equation}
S_n=\frac{\ln Z_n-n\ln Z_1}{1-n}. 
\label{RE}
\end{equation}
When $n\rightarrow 1$, we can extract EE from R\'enyi entropy. 
The $n$-sheet torus partition function is one-loop exact. 
Therefore, the contribution of the $n$-sheet partition function is up to the one-loop order \cite{Huang:2019nfm,Huang:2023aqz}. 
When applying the Weyl transformation from the cylinder manifold to a sphere manifold without the top and bottom points, the Liouville theory also appears and decouples from the 2D Schwarzian theory as in the torus case \cite{Huang:2023aqz}. 
The Liouville theory does not provide the backreaction to affect the classical solution \cite{Huang:2023aqz}. 
Therefore, the $n$-sheet partition function is a product of the classical $n$-sheet partition-function ($Z_{n, c}$) and the one-loop $n$-sheet partition-function ($Z_{n, q}$) 
\bea
Z_n=Z_{n, c}\cdot Z_{n, q}.
\eea 
The logarithm on the $n$-sheet partition function in calculating $S_n$,
\bea
\ln Z_n=\ln Z_{n, c}+\ln Z_{n, q}.
\eea
does not mix the classical and the one-loop terms. 
The boundary fields can be expanded as \cite{Cotler:2018zff}: 
\bea
\phi=\frac{\theta}{n}+\epsilon(y, \theta); \qquad \bar{\phi}=-\frac{\theta}{n}+\bar{\epsilon}(y, \theta),
\eea
where
\bea
\epsilon(y, \theta)&\equiv&\sum_{j_1, k_1}\epsilon_{j_1, k_1} e^{i\frac{j_1}{n}\theta-\frac{k_1}{\tau}y}; \qquad
\epsilon_{j_1, k_1}^*\equiv\epsilon_{-j_1, -k_1},
\nn\\
\bar{\epsilon}(y, \theta)&\equiv&\sum_{j_1, k_1}\bar{\epsilon}_{j_1, k_1} e^{i\frac{j_1}{n}\theta-\frac{k_1}{\tau}y}; \qquad
\bar{\epsilon}_{j_1, k_1}^*\equiv\bar{\epsilon}_{-j_1, -k_1}, 
\eea
where 
\bea
\tau_1=\tau.
\eea
The saddle-points are the $\theta/n$ and the $-\theta/n$ for the $\phi$ and $\bar{\phi}$, respectively \cite{Cotler:2018zff}. 
Each Fourier mode of the fluctuation, $\epsilon$ and $\bar{\epsilon}$, has three zero-modes: 
\bea
\epsilon_{j_1, k_1}=0; \qquad \bar{\epsilon}_{j_1, k_1}=0, \ j_1=-1, 0, 1
\eea 
due to the SL(2) gauge symmetry \cite{Cotler:2018zff}. 
\\ 

\noindent 
We first substitute the saddle points into the action. 
We then obtain 
\bea
\ln Z_{n, c}=\frac{c_{\mathrm{cft}_2}}{6n}\ln\bigg(\frac{L}{\epsilon}\bigg)
+n\frac{c_{\mathrm{cft}_2}}{6}\ln\bigg(\frac{L}{\epsilon}\bigg),
\eea
where the first term is from the 2D Schwarzian theory, and the second term is from the Liouville theory. 
Therefore, we derive the contribution of R\'enyi entropy from the saddle points: 
\bea
S_{n, c}=\frac{1}{1-n}\bigg\lbrack\frac{c_{\mathrm{cft}_2}}{6n}\ln\bigg(\frac{L}{\epsilon}\bigg)
-n\frac{c_{\mathrm{cft}_2}}{6}\ln\bigg(\frac{L}{\epsilon}\bigg)\bigg\rbrack
=\frac{c_{\mathrm{cft}_2}(1+n)}{6n}\ln\frac{L}{\epsilon}. 
\eea
When $n\rightarrow 1$, we reproduce the known result of CFT$_2$ 
\bea
\lim_{n\rightarrow 1} S_n=\frac{c_{\mathrm{cft}_2}}{3}\ln\frac{L}{\epsilon}.
\eea 
\\

\noindent
Now we discuss the one-loop contribution of $S_n$ from the $\epsilon(y, \theta)$ and $\bar{\epsilon}(y, \theta)$ \cite{Huang:2019nfm,Huang:2023aqz}. 
The expansion from the $\epsilon$ in the boundary action is \cite{Huang:2019nfm,Huang:2023aqz}
\bea
&&\frac{k}{4\pi}\int_{-\pi\mathrm{Im}(\tau)}^{\pi\mathrm{Im}(\tau)} dy\int_0^{2\pi n}d\theta\ \bigg(n^2\big(\partial_{\theta}^2\epsilon(y, \theta)\big)\big(\bar{\partial}\partial_{\theta}\epsilon(y, \theta)\big)
\nn\\
&&
-\big(\partial_{\theta}\epsilon(y, \theta)\big)\big(\bar{\partial}\epsilon(y, \theta)\big)\bigg)
\nn\\
&=&
-i\frac{k}{4\pi}n\tau\cdot\bigg\{ n^2\sum_{j_1, k_1}\bigg\lbrack\bigg(-\frac{j_1^2}{n^2}\cdot\frac{1}{2}\cdot\bigg(i\frac{k_1}{\tau}+i\frac{j_1}{n}\bigg)\bigg(i\frac{j_1}{n}\bigg)\bigg\rbrack|\epsilon_{j_1, k_1}|^2
\nn\\
&&
-\sum_{j_1 ,k_1}\bigg\lbrack\bigg(i\frac{j_1}{n}\bigg)\cdot\frac{1}{2}\cdot\bigg(i\frac{k_1}{\tau}+i\frac{j_1}{n}\bigg)\bigg\rbrack|\epsilon_{j_1, k_1}|^2\bigg\}
\nn\\
&=&-i\frac{k}{8\pi}\sum_{j_1, k_1}j_1(j_1^2-1)\bigg(k_1+\frac{j_1}{n}\tau\bigg)|\epsilon_{j_1, k_1}|^2,
\eea
where
\bea
\bar{\partial}\equiv\frac{1}{2}(-i\partial_y+\partial_{\theta}).
\eea
Taking the derivative of $\tau$ on the logarithm of the $n$-sheet one-loop partition gives 
\bea
\partial_{\tau}\ln Z_{n, q}=-\sum_{j_1\neq 0,\pm 1}\sum_{k_1=-\infty}^{\infty}\frac{\frac{j_1}{n}}{k_1+\frac{j_1}{n}\tau}.
\eea
The following fact of the digamma function
\bea
\tilde{\psi}(1-x)-\tilde{\psi}(x)=\pi\cot(\pi x),
\eea
in which the digamma function is defined by
\bea
\tilde{\psi}(a)\equiv
-\sum_{n_1=0}^{\infty}\frac{1}{n_1+a}, 
\eea
is helpful to simplify the complicated summation in the $n$-sheet partition function:
\bea
\sum_{m=-\infty}^{\infty}\frac{1}{m-x}&=&-\sum_{m=0}^{\infty}\frac{1}{m+x}+\sum_{m=1}^{\infty}\frac{1}{m-x}
\nn\\
&=&-\sum_{m=0}^{\infty}\frac{1}{m+x}+\sum_{m=0}^{\infty}\frac{1}{m+1-x}
=\tilde{\psi}(x)-\tilde{\psi}(1-x)
\nn\\
&=&-\pi\cdot\cot(\pi x).
\eea
Hence we obtain:
\bea
\partial_{\tau}\ln Z_{n, q}&=&-\sum_{j_1\neq 0,\pm 1}\sum_{k_1=-\infty}^{\infty}\frac{\frac{j_1}{n}}{k_1+\frac{j_1}{n}\tau}
=-\sum_{j_1\neq 0, \pm 1}\bigg(\frac{j_1}{n}\pi\bigg)\cdot\cot\bigg(\frac{j_1}{n}\pi\tau\bigg)
\nn\\
&=&-2\pi\sum_{j=2}^{\infty}\bigg(\frac{j_1}{n}\bigg)\cdot\cot\bigg(\frac{j_1\pi\tau}{n}\bigg).
\eea
\\

\noindent
To obtain a universal term, we regularize the series:
\bea
\partial_{\tau}\ln Z_{n, q}
&=&-2\pi\sum_{j_1=2}^{\infty}\bigg(\frac{j_1}{n}\bigg)\cdot\cot\bigg(\frac{j_1\pi\tau}{n}\bigg)
\nn\\
&=&-2\pi\sum_{j_1=2}^{\infty}\frac{j_1}{n}\cdot\bigg\lbrack\cot\bigg(\frac{j_1\pi\tau}{n}\bigg)+i\bigg\rbrack
+2\pi i\sum_{j_1=2}^{\infty}\frac{j_1}{n}.
\eea 
The series
\bea
\sum_{j_1=1}^{\infty} j_1=1+2\cdots
\eea
is divergent meaning that it does not converge to a finite value. 
We introduce the regularization to the summation: 
\bea
\lim_{\epsilon\rightarrow 0}\sum_{n=1}^{\infty} e^{-n \epsilon}n
&=&-\lim_{\epsilon\rightarrow 0}\frac{d}{d\epsilon}\sum_{n=1}^{\infty} e^{-n\epsilon}
=-\lim_{\epsilon\rightarrow 0}\frac{d}{d\epsilon}\frac{e^{-\epsilon}}{1-e^{-\epsilon}}
\nn\\
&=&-\lim_{\epsilon\rightarrow 0}\frac{d}{d\epsilon}\frac{1}{e^{\epsilon}-1}
=-\lim_{\epsilon\rightarrow 0}\frac{d}{d\epsilon}\bigg(\frac{1}{\epsilon}-\frac{1}{2}+\frac{\epsilon}{12}\bigg)
\nn\\
&=&-\frac{1}{12}+\lim_{\epsilon\rightarrow 0}\frac{1}{\epsilon^2}.
\eea
The fourth equality uses:
\bea
e^{\epsilon}-1
&=&\epsilon+\frac{\epsilon^2}{2}+\frac{\epsilon^3}{6}+\cdots, 
\nn\\
\frac{1}{e^{\epsilon}-1}
&=&\frac{1}{\epsilon}\frac{1}{1+\frac{\epsilon}{2}+\frac{\epsilon^2}{6}+\cdots}
=\frac{1}{\epsilon}\bigg(1-\frac{\epsilon}{2}-\frac{\epsilon^2}{6}+\frac{\epsilon^2}{4}+\cdots\bigg)
\nn\\
&=&\frac{1}{\epsilon}-\frac{1}{2}+\frac{\epsilon}{12}+\cdots. 
\nn\\
\eea
We can apply the following result
\bea
\sum_{j_1=1}^{\infty}j_1\rightarrow -\frac{1}{12}.
\eea
to obtain:
\bea
\partial_{\tau}\ln Z_{n, q}&=&-2\pi\sum_{j_1=2}^{\infty}\frac{j_1}{n}\cdot\bigg\lbrack\cot\bigg(\frac{j_1\pi\tau}{n}\bigg)+i\bigg\rbrack
+2\pi i\sum_{j_1=2}^{\infty}\frac{j_1}{n}
\nn\\
&\rightarrow&-2\pi\sum_{j_1=2}^{\infty}\frac{j_1}{n}\cdot\bigg\lbrack\cot\bigg(\frac{j_1\pi\tau}{n}\bigg)+i\bigg\rbrack
-i\frac{13\pi }{6n}.
\eea
\\

\noindent
We integrate the $\tau$, we obtain
\bea
\ln Z_{n, q}=-2\sum_{j_1=2}^{\infty}\bigg\lbrack\ln\sin\bigg(\frac{\pi j_1\tau}{n}\bigg)+i\frac{j_1\pi\tau}{n}\bigg\rbrack-i\frac{13\pi\tau}{6n}+\cdots,
\eea
where $\cdots$ is independent of the $\tau$. 
Because the series is convergent in $\ln Z_{n, q}$  when $\mathrm{Im}(\tau)>0$,  we obtain
\bea
\ln Z_{n, q}=\frac{13}{3n}\ln\frac{L}{\epsilon}
\eea
when considering the limit 
\bea
\frac{L}{\epsilon}\rightarrow\infty. 
\eea
Hence we obtain
\bea
\ln Z_{n, q}=\frac{13}{6n}\ln\frac{L}{\epsilon},
\eea
and the Rényi entropy for the one-loop correction is  \cite{Huang:2019nfm,Huang:2023aqz}:
\bea
S_{n, q}=\frac{1}{1-n}\big(\ln Z_{n, q}-n\ln Z_{1, q}\big)=\frac{13(n+1)}{6n}\ln\frac{L}{\epsilon}.
\eea
This result implies that the quantum correction only shifts the value of the central charge 
\bea
C_{\mathrm{cft}_2}=c_{\mathrm{cft}_2}+13
\eea
in the Rényi entropy \cite{Huang:2019nfm,Huang:2023aqz}
\bea
S_n=S_{n,c}+S_{n, q}
 \eea
 and EE \cite{Huang:2019nfm,Huang:2023aqz}
 \bea
 \lim_{n\rightarrow 1}S_n. 
 \eea
Hence the large central charge limit of CFT$_2$ includes the classical and one-loop contributions from AdS$_3$ bulk gravity. 
The result should be due to the different perturbation parameters between the bulk gravity and CFT. 

\subsection{Minimum Surface=Entanglement Entropy}
\noindent 
We introduce Wilson line to the action for the $n$-sheet case \cite{Ammon:2013hba,deBoer:2013vca}
\bea
&&
W_{\cal R}(C)
\nn\\
&=&\int DU DP\ \exp\bigg\lbrack\int_C ds\ \bigg(\mathrm{Tr}(PU^{-1}D_s U)+\lambda(s)\big(\mathrm{Tr}(P^2)-c_2\big)\bigg)\bigg\rbrack,
\eea
where
\bea
\sqrt{2c_2}=\frac{C_{cft_2}}{6}(1-n). 
\eea
The $U$ is an SL(2) element, $P$ is its conjugate momentum, and the covariant derivative is defined as the following:
\bea
D_sU\equiv \frac{d}{ds}U+A_sU-U\bar{A}_s, \qquad A_s\equiv A_{\mu}\frac{dx^{\mu}}{ds}, \qquad 
\bar{A}_s\equiv\bar{A}_{\mu}\frac{dx^{\mu}}{ds}.
\eea 
The endpoints of the Wilson line lie on the entangling surface of a single interval. 
We show the equation of motion of the $P$ \cite{Ammon:2013hba,deBoer:2013vca}
\bea
U^{-1}D_sU=-2\lambda P; 
\eea
the equation of motion of the $U$ \cite{Ammon:2013hba,deBoer:2013vca}
\bea
\frac{dP}{ds}=0; 
\eea
the equation of motion of the $\lambda(s)$ \cite{Ammon:2013hba,deBoer:2013vca}
\bea
\mathrm{Tr}P^2=c_2; 
\eea
the equation of the gauge field $A$ \cite{Ammon:2013hba,deBoer:2013vca}
\bea
i\frac{k}{2\pi}F_{\mu\nu}=-\int ds\ \frac{dx^{\rho}}{ds}\epsilon_{\mu\nu\rho}\delta^3\big(x-x(s)\big)UPU^{-1}; 
\eea
the equation of the gauge field $\bar{A}$ \cite{Ammon:2013hba,deBoer:2013vca}
\bea
i\frac{k}{2\pi}\bar{F}_{\mu\nu}=-\int ds\ \frac{dx^{\rho}}{ds}\epsilon_{\mu\nu\rho}\delta^3\big(x-x(s)\big)P.
\eea
\\

\noindent
The solution can be written as follows \cite{Ammon:2013hba,deBoer:2013vca}:
\bea
A&=&g^{-1}ag+g^{-1}dg, \qquad g=\exp(L_1 z_y)\exp(\rho L_0);
\nn\\
\bar{A}&=&\bar{g}^{-1}a\bar{g}^{-1}+\bar{g}^{-1}d\bar{g}, \qquad \bar{g}=\exp(L_{-1}\bar{z}_y)\exp(-\rho L_0),
\eea
where the gauge field is defined by that
\bea
a\equiv\sqrt{\frac{c_2}{2}}\frac{1}{k}\bigg(\frac{dz_y}{z_y}-\frac{d\bar{z}_y}{\bar{z}_y}\bigg)L_0, 
\eea 
corresponding to the choice: 
\bea
\rho(s)=s, \qquad U(s)=1, \qquad P(s)=\sqrt{2c_2}L_0.
\eea 
$\rho$ is a variable denoting the radial direction. 
The new coordinate variables are given by: 
\bea
z_y=\theta+iy, \qquad \bar{z}_y=\theta-iy.
\eea 
The gauge field $a$ provides the holonomy \cite{Ammon:2013hba,deBoer:2013vca}
\bea
\oint a=2\pi i\frac{\sqrt{2 c_2}}{k}L_0.
\eea
We introduce a more convenient basis to the SL(2) algebras, $L_{-1}$; $L_0$; $L_1$, \cite{Ammon:2013hba,deBoer:2013vca}
\bea
\lbrack L_m, L_n\rbrack=(m-n)L_{m+n}, \qquad m,n=0,\pm 1,
\eea
\bea
\mathrm{Tr}(L_0^2)=\frac{1}{2}, \qquad \mathrm{Tr}(L_{-1}L_1)=-1.
\eea 
The trace of other bilinears is zero. 
\\

\noindent
When we choose $c_2=0$ or $n=1$, the solution of the gauge fields is \cite{Ammon:2013hba,deBoer:2013vca}:
\bea
&&A
\nn\\
&=&g^{-1}dg
\nn\\
&=&\exp(-\rho L_0)\exp(-L_1z_y)L_1\exp(L_1z_y)\exp(\rho L_0)dz_y
\nn\\
&&
+\exp(-\rho L_0)\exp(-L_1z_y)\exp(L_1 z_y)L_0\exp(\rho L_0)d\rho
\nn\\
&=&
\exp(-\rho L_0)L_1\exp(\rho L_0)dz_y+L_0d\rho
\nn\\
&=&\exp(\rho)L_1dz_y+L_0d\rho, 
\nn\\
&&\bar{A}
\nn\\
&=&\bar{g}^{-1}d\bar{g}
\nn\\
&=&\exp(\rho L_0)\exp(-L_{-1}\bar{z}_y)L_{-1}\exp(L_{-1}\bar{z}_y)\exp(-\rho L_0)d\bar{z}_y
\nn\\
&&
-\exp(\rho L_0)\exp(-L_{-1}\bar{z}_y)\exp(L_{-1}\bar{z}_y)L_0\exp(-\rho L_0)d\rho
\nn\\
&=&
\exp(\rho L_0)L_{-1}\exp(-\rho L_0)d\bar{z}_y-L_0d\rho
\nn\\
&=&\exp(\rho)L_{-1}d\bar{z}_y-L_0d\rho.
\eea
Therefore, the vielbein becomes \cite{Ammon:2013hba,deBoer:2013vca}:
\bea
e=\frac{1}{2}(A-\bar{A})=-\frac{1}{2}\exp(\rho)L_1dz_y+\frac{1}{2}\exp(\rho)L_{-1}d\bar{z}_y+L_0d\rho.
\eea
\\

\noindent
We can use the vielbein to determine the metric \cite{Ammon:2013hba,deBoer:2013vca}
\bea
g_{\mu\nu}=2\mathrm{Tr}(e_{\mu}e_{\nu}). 
\eea
The spacetime interval is \cite{Ammon:2013hba,deBoer:2013vca}
\bea
ds^2=d\rho^2+e^{2\rho}dz_yd\bar{z}_y.
\eea
We then redefine $z_y$ and $\bar{z}_y$: 
\bea
z_y=re^{i\Phi}, \qquad \bar{z}_y=re^{-i\Phi}, 
\eea
where 
\bea
0<r<\infty;  0 \ge\Phi<2\pi, 
\eea
to get the spacetime interval \cite{Ammon:2013hba,deBoer:2013vca}
\bea
ds^2=d\rho^2+e^{2\rho}(dr^2+r^2d\Phi^2).
\eea 
We then redefine the coordinate variables as 
\bea
\tilde{z}\equiv e^{-\rho}. 
\eea
$ds^2$ is precisely the AdS$_3$ solution in the Poincaré coordinate with the Euclidean signature \cite{Ammon:2013hba,deBoer:2013vca}
\bea
ds^2=\frac{1}{\tilde{z}^2}(d\tilde{z}^2+dr^2+\tilde{r}^2d\Phi^2). 
\eea 
When $n$ approaches 1, the geometry becomes the AdS$_3$ manifold. 
\\

\begin{figure}[!htb]
\begin{center}
\hspace*{-1.3cm}
\includegraphics[width=1.\textwidth]{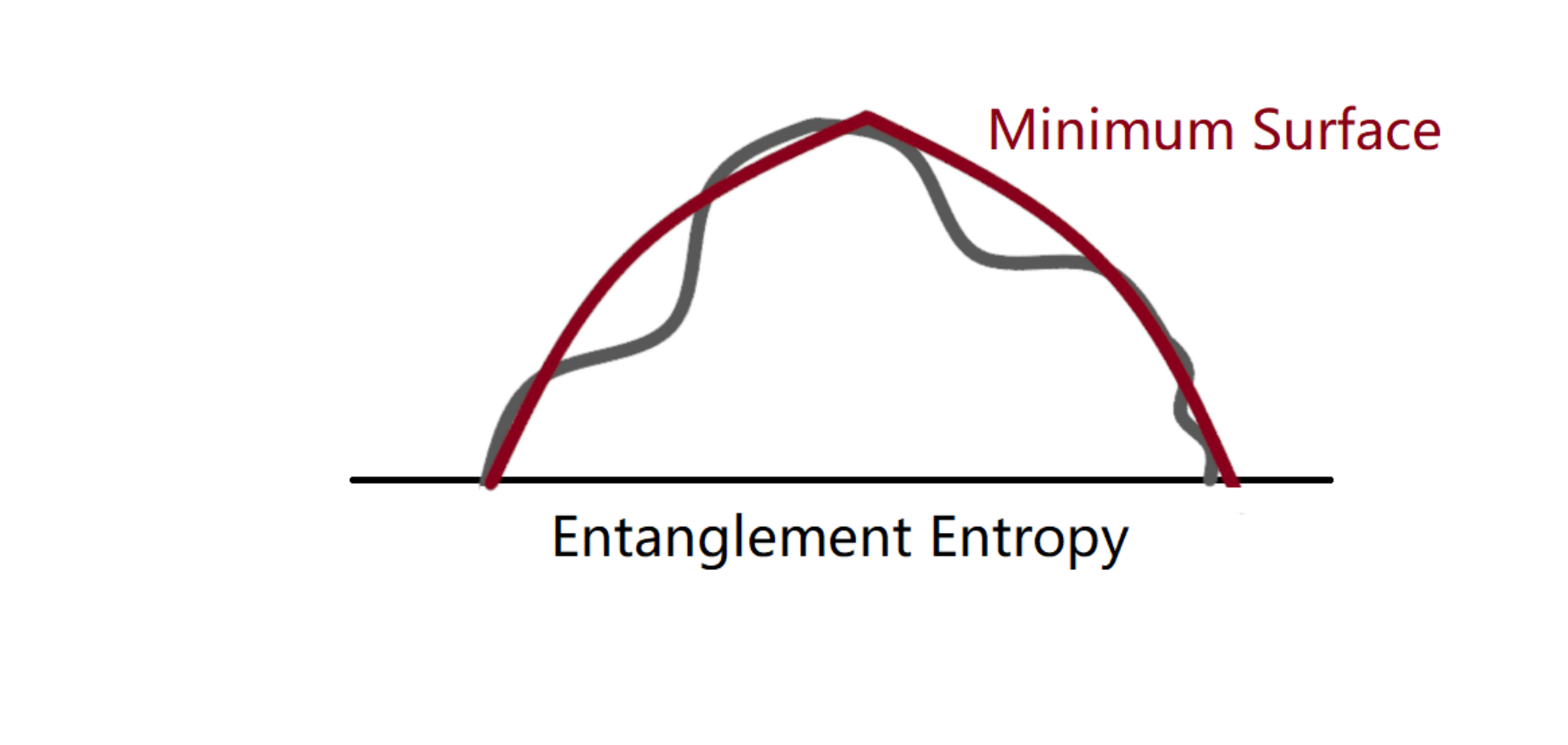}
\caption{The minimum surface gives the on-shell contribution to the boundary theory's EE. 
The gauge fields cause surface fluctuations, which deform the central charge in EE. 
}
\label{MSEE.pdf}
\end{center}
\end{figure}

\noindent 
Now we introduce the non-trivial gauge field $a$ to consider the backreaction \cite{Ammon:2013hba,deBoer:2013vca}:
\bea
A&=&\exp(\rho)L_1 dz_y+L_0d\rho+\frac{1}{2k}\sqrt{\frac{c_2}{2}}(L_0-z_ye^{\rho}L_1)\bigg(\frac{dz_y}{z_y}-\frac{d\bar{z}_y}{\bar{z}_y}\bigg),
\nn\\
\bar{A}&=&\exp(\rho)L_{-1}d\bar{z}_y-L_0d\rho+\frac{1}{2k}\sqrt{\frac{c_2}{2}}(L_0+\bar{z}_ye^{\rho}L_{-1})\bigg(\frac{dz_y}{z_y}-\frac{d\bar{z}_y}{\bar{z}_y}\bigg).
\eea
The vielbein is \cite{Ammon:2013hba,deBoer:2013vca}:
\bea
&&e
\nn\\
&=&
\frac{1}{2}(A-\bar{A})
\nn\\
&=&
\frac{1}{2}\exp(\rho)L_1dz_y-\frac{1}{2}\exp(\rho)L_{-1}d\bar{z}_y+L_0d\rho
\nn\\
&&
+\frac{1}{2k}\sqrt{\frac{c_2}{2}}(L_0-z_ye^{\rho}L_1)\bigg(\frac{dz_y}{z_y}-\frac{d\bar{z}_y}{\bar{z}_y}\bigg)
\nn\\
&&
-\frac{1}{2k}\sqrt{\frac{c_2}{2}}(L_0+\bar{z}_ye^{\rho}L_{-1})\bigg(\frac{dz_y}{z_y}-\frac{d\bar{z}_y}{\bar{z}_y}\bigg).
\nn\\
\eea
The backreaction provides the $n$-sheet geometry \cite{Ammon:2013hba,deBoer:2013vca}: 
\bea
&&ds^2
\nn\\
&=&
d\rho^2+e^{2\rho}dz_yd\bar{z}_y
\nn\\
&&-\frac{1}{k}\sqrt{\frac{c_2}{2}}e^{2\rho}\bar{z}_ydz_y(2id\Phi)\mathrm{Tr}(L_1L_{-1})
+\frac{1}{k}\sqrt{\frac{c_2}{2}}z_yd\bar{z}_y(2id\Phi)\mathrm{Tr}(L_{-1}L_1)
\nn\\
&&
+\frac{c_2}{2k^2}e^{2\rho}z_y\bar{z}_y(-4d\Phi^2)\mathrm{Tr}(L_1L_{-1})
\nn\\
&=&
d\rho^2+e^{2\rho}dz_yd\bar{z}_y
+\frac{1}{k}\sqrt{\frac{c_2}{2}}e^{2\rho}(2i r^2d\Phi)(2id\Phi)
+\frac{2c_2}{k^2}e^{2\rho}r^2d\Phi^2
\nn\\
&=&
d\rho^2+e^{2\rho}(dr^2+r^2d\Phi^2)-e^{2\rho}\frac{2\sqrt{2c_2}}{k}r^2d\Phi^2+
e^{2\rho}\frac{2c_2}{k^2}r^2d\Phi^2
\nn\\
&=&d\rho^2+e^{2\rho}\bigg\lbrack dr^2+\bigg(\frac{\sqrt{2c_2}}{k}-1\bigg)^2r^2d\Phi^2\bigg\rbrack
\nn\\
&=&
d\rho^2+e^{2\rho}(dr^2+n^2r^2d\Phi^2).
\eea
At the boundary $\rho\rightarrow\infty$, we obtain the $n$-sheet cylinder 
\bea
ds_b^2=d\tilde{y}^2+n^2d\Phi^2, 
\eea
by using 
\bea
r\equiv e^{\tilde{y}}
\eea
and omitting the pre-factor. 
The pre-factor generates the Liouville theory to compensate for the difference between the cylinder and sphere partition functions. 
Because the Wilson line $W_{\cal R}$ does not survive under the limit $n\rightarrow 1$, the boundary action only leaves the 2D Schwarzian theory and the Liouville theory \cite{Huang:2019nfm,Huang:2023aqz}. 
Hence computing the Wilson line in the AdS$_3$ Einstein gravity theory is equivalent to computing $Z_n/Z^n_1$ in 2D Schwarzian theory and Liouville theory \cite{Huang:2019nfm,Huang:2023aqz}
 \bea
 \langle W_{\cal R}\rangle=\frac{Z_n}{Z_1^n}+{\cal O}(n-1), 
 \eea 
 where $\langle W_{\cal R}\rangle$ is the expectation value of the Wilson line.
 The expression for EE can be restated using the Wilson line \cite{Huang:2019nfm,Huang:2023aqz}
\bea
S_{EE}=\lim_{n\rightarrow 1}\frac{1}{1-n}\ln\langle W_{\cal R}\rangle,
\eea
We conclude that the Wilson line should be a suitable operator playing the role of minimum surface (Fig. \ref{MSEE.pdf}) \cite{Huang:2019nfm,Huang:2023aqz}. 

\section{Outlook and Future Directions}
\label{sec:5}
\noindent 
We discussed various topics related to the context of AdS$_3$ Einstein gravity and its connection to CFT. 
AdS$_3$ pure Einstein gravity is a crucial tool for exploring the relationship between gravity and field theory. 
However, Einstein gravity with asymptotic AdS$_3$ boundary conditions is not considered a physical theory because it lacks a unitary dual CFT on the boundary. 
This was pointed out by Ref. \cite{Witten:2007kt}. 
A physical theory is unnecessarily needed to address the duality between the bulk and the boundary. 
3D Einstein gravity theory can be made computable through renormalization techniques \cite{Witten:1988hc}, which help deal with divergences. 
The Weyl transformation leads to the Liouville theory in 2D Schwarzian theory, which is not a conventional CFT$_2$. 
The Liouville theory simplifies computations in the context of AdS/CFT correspondence \cite{Huang:2023aqz}. 
It is desirable to have the exact mapping on the bulk and boundary sides. 
String Theory suggests that introducing matter fields can lead to a dual unitary CFT.  
However, these matter fields can produce backreaction in the bulk, making the correspondence more complex. 
Given the close connection between 3D Einstein gravity and 2D dilaton gravity, we suggest starting with 2D gravity exploration to understand aspects of the correspondence better. 
We are interested in exploring the intricacies of the AdS/CFT correspondence, including the challenges related to unitarity, backreaction, and the role of matter fields. 
Theoretical physics is a continuously evolving field of research, and one area of particular interest is the holographic nature of the correspondence. 
Researchers are working hard to find explicit examples and mathematical frameworks to better understand these phenomena. 
By doing so, they hope to uncover new insights into the fundamental nature of gravity and quantum field theory.
\\

\noindent 
We discuss the relationship between central charges in CFT$_2$ and the bare gravitation constant in 3D Einstein gravity ($G_3$). 
In CFT$_2$, the central charge is an important parameter that characterizes the algebraic and geometric properties of the theory. 
It is not directly proportional to the inverse of $G_3$, which suggests that these two quantities are not simply related in this context \cite{Cotler:2018zff}. 
The inverse central charge order corresponds to an infinite number of terms in the perturbation expansion of $G_3$.  
This means that when attempting to describe 3D Einstein gravity as a perturbative theory, there are an infinite number of terms that must be considered. 
These terms are captured by a proportional relationship to the inverse central charge order. 
It suggests that CFT$_2$ serves as a framework for a perturbative resummation of gravity, which can yield non-perturbative results. 
To avoid the non-renormalizability in the higher-dimensional gravity, the resummation of infinite expansion terms or the non-locality should be a simple solution. 
This is an interesting idea, as it hints that studying Quantum Gravity within the context of CFT$_2$ might provide insights that go beyond traditional perturbative approaches.  
CFT should be a better approach to studying Quantum Gravity. 
This could simplify the description of certain aspects of Quantum Gravity. 
We have established an interesting link between CFT$_2$ and AdS$_3$ Einstein gravity. 
Our findings indicate that the central charge in CFT$_2$ and the bare gravitational constant $G_3$ are not inversely proportional as previously thought \cite{Cotler:2018zff}. 
However, we expect that there is a deeper connection that deserves further exploration as it could offer valuable insights into the field of Quantum Gravity.
\\

\noindent 
We are currently discussing some key concepts at the intersection of Quantum Information and Holography within the context of Theoretical Physics and Quantum Gravity \cite{Ryu:2006bv}. 
The concepts from quantum information theory can be applied to classical gravity systems to analyze them. 
This analysis can potentially reveal new insights into emergent phenomena, as suggested by  Ref. \cite{Ryu:2006bv}. 
Quantum Information Measures can help in comprehending the emergence of spacetime in certain physical systems. 
This connection is intriguing because it indicates that the geometry of spacetime might be fundamentally linked to the quantum entanglement structure of the underlying degrees of freedom. 
The study of holography and related topics can benefit from Quantum Information Measures as they can sometimes lead to simple and exact results, as shown by Refs. \cite{Huang:2019nfm,Huang:2023aqz}. 
Quantum information considerations possibly help resolve or shed light on the black hole information paradox. 
The paradox arises from apparent conflicts between the principles of quantum mechanics and general relativity in the context of black holes. 
Quantum Information Measures could offer novel approaches to comprehend black holes and spacetime beyond traditional methods. 
Researching Quantum Gravity from the perspective of Quantum Information is a promising and fascinating avenue. 
This approach possibly provides fresh insights into the nature of gravity and spacetime at the quantum level. 
We highlight the growing importance of Quantum Information Measures in the study of holography, quantum gravity, and the black hole information paradox. 
It underscores the potential for quantum information theory to provide new tools and perspectives for addressing fundamental questions in theoretical physics. 

\section*{Acknowledgments}
\noindent
The author thanks Chuan-Tsung Chan, Bartlomiej Czech, Jan de Boer, Xing Huang, Kristan Jensen, Hongfei Shu, Ryo Suzuki, and Chih-Hung Wu for their discussion and would thank Nan-Peng Ma for his encouragement.
CTM acknowledges the Nuclear Physics Quantum Horizons program through the Early Career Award (Grant No. DE-SC0021892); 
YST Program of the APCTP;  
Post-Doctoral International Exchange Program; 
China Postdoctoral Science Foundation, Postdoctoral General Funding: Second Class (Grant No. 2019M652926); 
Foreign Young Talents Program (Grant No. QN20200230017). 
The author thanks the National Tsing Hua University, the Institute for Advanced Study at the Tsinghua University, and the Center for Quantum Science at the Sogang University. 
Discussion during the workshops, ``East Asia Joint Workshop on Fields and Strings 2019'' and ``The 17th Italian-Korean Symposium for Relativistic Astrophysics'', was helpful to this work. 


\end{document}